\documentclass[a4paper,12pt]{article}  
\pdfoutput=1
\usepackage{hyperref}
\usepackage{epsfig}
\usepackage{amssymb}
\usepackage{cancel}
\usepackage{setspace}
\usepackage[usenames,dvipsnames,svgnames,table]{xcolor}
\usepackage{amsmath}
\usepackage{graphicx}
\usepackage{subfigure}
\usepackage{url}
\usepackage{slashed}
\usepackage{booktabs}
\usepackage{graphbox}
\usepackage{placeins} 
\usepackage{cite}

\newcommand{\be}{\begin{equation}}
\newcommand{\ee}{\end{equation}}
\newcommand{\br}{\begin{eqnarray}}
\newcommand{\bea}{\begin{eqnarray}}
\newcommand{\eea}{\end{eqnarray}}
\newcommand{\er}{\end{eqnarray}}
\newcommand{\ba}{\begin{array}}
\newcommand{\ea}{\end{array}}
\newcommand{\bi}{\begin{itemize}}
\newcommand{\ei}{\end{itemize}}
\newcommand{\bn}{\begin{enumerate}}
\newcommand{\en}{\end{enumerate}}
\newcommand{\bc}{\begin{center}}
\newcommand{\ec}{\end{center}}

\newcommand{\beq}{\begin{equation}}
\newcommand{\eeq}{\end{equation}}

\newcommand{\U}{\scriptscriptstyle U}
\newcommand{\D}{\scriptscriptstyle D}

\newcommand{\E}{\scriptscriptstyle E}
\newcommand{\N}{\scriptscriptstyle N}
\newcommand{\LL}{\scriptscriptstyle L}
\newcommand{\RR}{\scriptscriptstyle R}
\newcommand{\LR}{\scriptscriptstyle LR}
\newcommand{\X}{\scriptscriptstyle X}
\newcommand{\Y}{\scriptscriptstyle Y}

\newcommand{\gsim}{\lower1.0ex\hbox{$\;\stackrel{\textstyle>}{\sim}\;$}}
\newcommand{\lsim}{\lower1.0ex\hbox{$\;\stackrel{\textstyle<}{\sim}\;$}}

\newcommand{\bs}{\begin{small}}
\newcommand{\es}{\end{small}}

\newcommand{\gUL}{g^{\scriptscriptstyle{U}}_{\scriptscriptstyle{L}}}
\newcommand{\gUR}{g^{\scriptscriptstyle{U}}_{\scriptscriptstyle{R}}}
\newcommand{\gDL}{g^{\scriptscriptstyle{D}}_{\scriptscriptstyle{L}}}
\newcommand{\gDR}{g^{\scriptscriptstyle{D}}_{\scriptscriptstyle{R}}}
\newcommand{\gEL}{g^{\scriptscriptstyle{E}}_{\scriptscriptstyle{L}}}
\newcommand{\gER}{g^{\scriptscriptstyle{E}}_{\scriptscriptstyle{R}}}
\newcommand{\gNL}{g^{\scriptscriptstyle{N}}_{\scriptscriptstyle{L}}}
\newcommand{\gNR}{g^{\scriptscriptstyle{N}}_{\scriptscriptstyle{R}}}

\newcommand{\lamSU}{\lambda^{\scriptscriptstyle{U}}_{\scriptscriptstyle{S}}}
\newcommand{\lamSD}{\lambda^{\scriptscriptstyle{D}}_{\scriptscriptstyle{S}}}
\newcommand{\lamSN}{\lambda^{\scriptscriptstyle{N}}_{\scriptscriptstyle{S}}}
\newcommand{\lamSL}{\lambda^{\scriptscriptstyle{E}}_{\scriptscriptstyle{S}}}

\newcommand{\lamSX}{\lambda^{\scriptscriptstyle{X}}_{\scriptscriptstyle{S}}}

\newcommand{\lamULL}{\lambda^{\scriptscriptstyle{U}}_{\scriptscriptstyle{LL}}}
\newcommand{\lamDLL}{\lambda^{\scriptscriptstyle{D}}_{\scriptscriptstyle{LL}}}
\newcommand{\lamURR}{\lambda^{\scriptscriptstyle{U}}_{\scriptscriptstyle{RR}}}
\newcommand{\lamDRR}{\lambda^{\scriptscriptstyle{D}}_{\scriptscriptstyle{RR}}}

\newcommand{\lamULR}{\lambda^{\scriptscriptstyle{U}}_{\scriptscriptstyle{LR}}}
\newcommand{\lamDLR}{\lambda^{\scriptscriptstyle{D}}_{\scriptscriptstyle{LR}}}

\newcommand{\lamNLL}{\lambda^{\scriptscriptstyle{N}}_{\scriptscriptstyle{LL}}}
\newcommand{\lamELL}{\lambda^{\scriptscriptstyle{E}}_{\scriptscriptstyle{LL}}}
\newcommand{\lamNRR}{\lambda^{\scriptscriptstyle{N}}_{\scriptscriptstyle{RR}}}
\newcommand{\lamERR}{\lambda^{\scriptscriptstyle{E}}_{\scriptscriptstyle{RR}}}

\newcommand{\lamNLR}{\lambda^{\scriptscriptstyle{N}}_{\scriptscriptstyle{LR}}}
\newcommand{\lamELR}{\lambda^{\scriptscriptstyle{E}}_{\scriptscriptstyle{LR}}}

\newcommand{\lamHHS}{\lambda_{HH_S}}

\newcommand{\gD}{g_{\scriptscriptstyle{D}}}

\newcommand{\lamLLX}{\lambda^{\scriptscriptstyle{X}}_{\scriptscriptstyle{LL}}}
\newcommand{\lamRRX}{\lambda^{\scriptscriptstyle{X}}_{\scriptscriptstyle{RR}}}
\newcommand{\lamLRX}{\lambda^{\scriptscriptstyle{X}}_{\scriptscriptstyle{LR}}}

\newcommand{\LambdaUV}{\Lambda_{\rm UV}}

\newcommand{\eed}{e_{\scriptscriptstyle{D}}}

\newcommand{\lamLLQ}{\lambda^{q}_{\scriptscriptstyle{LL}}}
\newcommand{\lamLLL}{\lambda^{\scriptscriptstyle{L}}_{\scriptscriptstyle{LL}}}
\newcommand{\lamRRQ}{\lambda^{q}_{\scriptscriptstyle{RR}}}
\newcommand{\lamRRL}{\lambda^{\scriptscriptstyle{L}}_{\scriptscriptstyle{RR}}}
\newcommand{\lamLRl}{\lambda^{\scriptscriptstyle{L}}_{\scriptscriptstyle{LR}}}
\newcommand{\lamLRq}{\lambda^{q}_{\scriptscriptstyle{LR}}}
\newcommand{\lamSq}{\lambda^{q}_{\scriptscriptstyle{S}}}
\newcommand{\lamSl}{\lambda^{\scriptscriptstyle{L}}_{\scriptscriptstyle{S}}}
\newcommand{\gLL}{g^{\scriptscriptstyle{L}}_{\scriptscriptstyle{L}}}
\newcommand{\gLR}{g^{\scriptscriptstyle{L}}_{\scriptscriptstyle{R}}}
\newcommand{\gQL}{g^{q}_{\scriptscriptstyle{L}}}
\newcommand{\gQR}{g^{q}_{\scriptscriptstyle{R}}}
\newcommand{\SSS}{\scriptscriptstyle{S}}

\begin{document}
\begin{center}
{\Large {\bf Vacuum stability with radiative Yukawa couplings
}}
\\
\vspace*{1.5cm}
{
\large     

 { Emidio Gabrielli$^{a,b,c}$}, { Luca Marzola$^{c}$}, { Kristjan M\"u\"ursepp$^{c}$,   Ruiwen Ouyang$^{c}$}
}\\
 \vspace{0.5cm}
 {\it
 (a)  Dipartimento di Fisica, Theoretical section, Universit\`a di 
 Trieste, \\ Strada Costiera 11, I-34151 Trieste, Italy \\ 
 (b) INFN, Sezione di Trieste, Via Valerio 2, I-34127 Trieste, Italy
 \\
 (c) NICPB, R\"avala 10, Tallinn 10143, Estonia 
\\[1mm] }
\vspace*{2cm}{\bf ABSTRACT}
\end{center}
\vspace{0.3cm}

\noindent
We explore the electroweak vacuum stability in the framework of a recently proposed paradigm for the origin of Yukawa couplings. These arise as low energy effective couplings radiatively generated by portal interactions with a hidden, or dark, sector at the one-loop level. Possible tree-level Yukawa couplings are forbidden by a new underlying symmetry, assumed to be spontaneously broken by the vacuum expectation value of a new scalar field above the electroweak scale. As a consequence, the top Yukawa interaction ceases to behave as a local operator at energies above the new sector scale and, therefore, cannot contribute to the running of the quartic Higgs coupling  at higher energies. By studying two complementary scenarios, we explicitly show that the framework can achieve the stability of the electroweak vacuum without particular tuning of parameters. The proposed mechanism requires the existence of a dark sector and new portal messenger scalar interactions that, connecting the Standard Model to the dark sector fields, could be tested at the LHC and future collider experiments.

\newpage
\section{Introduction} 
\label{sec:introduction}
The discovery of the Higgs boson by the ATLAS and CMS collaborations in  proton-proton collisions data with centre-of-mass energies of 7 TeV and 8 TeV~\cite{Aad:2012tfa} has marked a milestone in our understanding of the electroweak symmetry breaking (EWSB). To-date, the tested properties of this particle are all in good agreement with the Standard Model (SM) predictions \cite{ATLAS:2019slw,Sirunyan:2018koj}. In particular, the recent observations of the Higgs boson decay modes into bottom quark~\cite{Sirunyan:2018kst,Aad:2020vbr,Aaboud:2018zhk} and tau-lepton pairs~\cite{Sirunyan:2017khh, ATLAS:2018lur} are consistent with the SM Yukawa coupling strength and, therefore, support the existence of the corresponding interactions in Nature. Likewise, the detection of the Higgs boson production in association with top anti-top quark pairs \cite{Sirunyan:2018hoz, Aaboud:2018urx} matches the SM expectations and thus provides a direct confirmation of the existence of top quark Yukawa coupling. Given the Higgs boson vacuum expectation value (VEV) inferred from weak interactions, $v=246$ GeV, these results inevitably strengthen our confidence in the SM and in the Yukawa coupling origin of elementary fermion masses.

The precise measurement of the Higgs boson mass, $m_H=125.25\pm 0.17$~\cite{Zyla:2020zbs}, has also disclosed the expected value of the quartic Higgs boson coupling at the electroweak (EW) scale, which is about $\lambda_H(m_t)\sim 0.126$. This inference allows us to speculate on the stability of the EW vacuum by computing the relevant quantum corrections to the Higgs scalar potential~\cite{Cabibbo:1979ay,Hung:1979dn,Lindner:1985uk,Sher:1988mj,Schrempp:1996fb,Altarelli:1994rb}.
For large field values, $H\gg v$, the renormalization group (RG) improved effective potential is well approximated by the tree-level form with a running coupling 
\bea
V_{\rm eff}^{\rm tree}(H)&\simeq&\frac{\lambda_H(\mu_H)}{4}H^4\, ,
\label{Veff}
\eea
where the scale $\mu_H\sim H$ is of the order of the Higgs field value. Therefore, the problem of vacuum stability can be studied by simply analyzing the RG evolution of $\lambda_H$ and, at least in principle, solved by requiring that $\lambda_H(\mu)>0$ up to energies close to the Planck scale.

Impressive efforts have been dedicated to the computation of the higher-order corrections to the RG flow controlling the evolution of the Higgs quartic coupling~\cite{Bezrukov:2009db,Ellis:2009tp,EliasMiro:2011aa,Isidori:2007vm,
Mihaila:2012fm,Degrassi:2012ry,Chetyrkin:2012rz}.
The resulting relation that at low energy connects $\lambda_H(\mu)$ to the Fermi constant $G_F$ is
\bea
\lambda_H(\mu)&=&\frac{G_F m_H^2}{\sqrt{2}}+\Delta \lambda_H(\mu)\, ,
\eea
where $\Delta \lambda_H(\mu)$ contains finite threshold corrections that arise beyond the tree-level. These corrections are quite large and are the main source of uncertainty in the determination of the value of $\lambda_H$ on the considered energy span. Recently, the determination of the next-to-next-to-leading-order (NNLO) corrections to $\Delta \lambda_H(\mu)$, including the complete two-loop Yukawa-QCD contributions~\cite{Degrassi:2012ry,Chetyrkin:2012rz}, has allowed a reduction of the error in the determination of the Higgs mass of about $\pm 0.7$ GeV~\cite{Degrassi:2012ry}.

Given the current experimental values of the SM parameters that enter the RG evolution of the Higgs quartic coupling, these analyses reveal that  $\lambda_H$ becomes negative well below the Planck scale and that the EW vacuum is thus metastable. In particular, taking into account the theoretical and experimental uncertainties, ensuring the absolute stability of the EW vacuum up to the Planck scale requires $m_H> (129.4\pm 1.8)$ GeV. Equivalently, the SM vacuum stability is excluded at $2\sigma$ for $m_H< 126$ GeV~\cite{Degrassi:2012ry}. The result still critically depends on the value of the top quark mass, which is the dominant source of uncertainty in the determination of the Higgs mass, and affects the RG equations (RGE) of the Higgs quartic coupling through the negative contribution induced by the related Yukawa coupling. 

A possible way to ensure the stability of SM vacuum is to assume that all Yukawa couplings, including that of the top quark, are effective low energy parameters. Importantly, these are to be radiatively generated in absence of direct couplings of the Higgs boson to any fermion field, that is, by requiring a fundamentally \emph{fermiophobic} Higgs boson.  In fact, if the SM Yukawa couplings were to be generated through new fundamental interactions of the Higgs boson with fermion fields, the vacuum instability problem could still occur due to the corresponding -- and potentially sizeable -- new fermion-loop contributions. An explicit example is provided by models based on the universal seesaw mechanism~\cite{Davidson:1987mh,Rajpoot:1987fca,Berezhiani:1983hm,Chang:1986bp,Berezhiani:1991ds,Dev:2015vjd}, which generate the SM Yukawa couplings through fundamental interaction of the Higgs boson with new vector-like fermions that might still drive the scalar potential to negative values.    

Provided that new physics (NP) is below the SM instability scale, the fermiophobic Higgs condition can instead guarantee the stability of the EW vacuum. The underlying idea is straightforward: radiatively generated Yukawa operators cease to be local operators above the NP scale where they are generated. Then, due to the fermiophobic nature of the Higgs boson, the RGE of $\lambda_H$ at higher energies receives only the positive contributions of the {\it bosonic} degrees of freedom which thus enforce the vacuum stability. Clearly, in order to ensure that the Higgs quartic coupling remain positive throughout its complete RG evolution, the NP scale where the SM Yukawa operators are effectively generated must be below the SM instability scale. The latter therefore provides a theoretical upper bound on the NP scale required for the validity of the proposed solution. 

The fermiophobic Higgs condition can be naturally implemented by extending the theory to any local or global symmetry that forbids all SM Yukawa operators. For instance, the mechanism is straightforwardly embedded in the scenarios of Refs.~\cite{Gabrielli:2013jka,Gabrielli:2016vbb,Gabrielli:2019sjg} which were originally designed to solve the flavor hierarchy problem. In more detail, the tree-level Yukawa operators are forbidden by a new symmetry $S$: a discrete symmetry in Ref.~\cite{Gabrielli:2013jka} and a local $SU(2)_R$ extension of the SM gauge group in Ref~\cite{Gabrielli:2016vbb,Gabrielli:2019sjg}. In either case, the new symmetry is spontaneously broken by the vacuum expectation value $v_S$ of a dedicated scalar field which, thereby, allows for the emergence of the SM Yukawa operators. 

The framework predicts the existence of massive vector-like dark fermion fields -- heavy SM gauge-singlet replicas of the SM fermions -- and a set of scalar messenger fields that mediate the interactions between the SM and the dark sector. The messenger fields carry the same quantum numbers of squarks and sleptons of known supersymmetric models, in addition to a new $U(1)_D$ gauge charge under which the dark sector fields are also charged. The chiral symmetry breaking necessary for the Yukawa coupling generation is provided by the dark-fermion masses and communicated, at the 1-loop level, to the SM fields by the messengers. After the spontaneous breaking of the symmetry $S$, the  emerging Yukawa couplings are then necessarily proportional to the involved dark-fermion mass. A non-perturbative dynamics in the dark-sector, related to the $U(1)_D$ gauge symmetry, is responsible for the exponential spread of the dark fermion masses and, therefore, for the observed hierarchy of the SM Yukawa couplings~\cite{Gabrielli:2007cp}. The same framework also allows for the radiative origin of flavor mixing, modelled in the Cabibbo-Kobayashi-Maskawa matrix~\cite{Gabrielli:2019sjg}.

Adopting the simplest model delineated by the framework~\cite{Gabrielli:2013jka}, in the present paper we explore the EW vacuum stability in light of the fermiophobic Higgs mechanism. In particular, we compute the 1-loop contributions to the  $\beta$-function of the Higgs quartic coupling induced by the new degrees of freedom and analyse the conditions required for the positivity of this parameter on the whole of its RG evolution.

The paper is organized as follows. In the next Section we discuss the theoretical framework at the basis of the construction, detailing the relevant interactions of the new fields. In Section~\ref{sec:3} we review how the SM Yukawa coupling are generated from the interactions of messengers and dark fermions, whereas in Section~\ref{sec:4} we compute the 1-loop $\beta$-functions of the model. In Section~\ref{sec:5} we study the vacuum stability of the theory by analyzing the RG evolution of the $\lambda_H$ coupling from the EW scale up to the ultraviolet (UV) cutoff of the theory. We conclude with Section~\ref{sec:6}, where we summarize our findings.

\section{Theoretical framework} 
\label{sec:Radiative Yukawa couplings}

We summarize here the main features of the model at the basis of the present work, using the original formulation of Ref.~\cite{Gabrielli:2013jka} for the sake of simplicity. Because the discussion of the vacuum stability issue does not significantly depend on the nature of the symmetry used to forbid the existence of tree-level Higgs Yukawa couplings, our results can be applied also to the framework based on the Left-Right (LR) gauge symmetry presented in Ref.~\cite{Gabrielli:2016vbb}.

\begin{table}[htb!]
  \centering
	\begin{tabular}{c|c|c|c|c|c|c}
		\toprule
		\textbf{Field} & \textbf{Spin} & $\mathbf{\mathbb{Z}_2}$ charge & $\mathbf{U(1)_{D}}$ charge& $\mathbf{U(1)_{Y}}$ charge& $\mathbf{SU(2)_L}$ repr.&  $\mathbf{SU(3)_c}$ repr. \\
		\midrule
		\rowcolor[gray]{.85}
		\multicolumn{7}{c}{\emph{Extended SM sector:}} \\
$q^i_L$  & 1/2 & 1 & 0 & 1/6 & 2 & 3 \\
$U^i_R$  & 1/2 & 1 & 0 & 2/3 & 1 & 3 \\
$D^i_R$  & 1/2 & 1 & 0 & -1/3 & 1 & 3 \\
\hline
$L^i_L$  & 1/2 & 1 & 0 & -1/2 & 2 & 1 \\
$E^i_R$  & 1/2 & 1 & 0 & -1 & 1 & 1 \\
$\nu^i_R$  & 1/2 & 1 & 0 & 0 & 1 & 1 \\
\hline
$\hat H$ & 0 & -1 & 0 & 1/2 & 2 & 1 \\
$H_S$  & 0 & -1 & 0 & 0 & 1 & 1 \\
		\rowcolor[gray]{.85}
		\multicolumn{7}{c}{\emph{Mediator sector:}}  \\
$\hat S^{\U_i}_L$ & 0 & 1 & $-\eed^{\U_i}$ & 1/6 & 2 & 3 \\
$\hat S^{\D_i}_L$ & 0 & 1 & $-\eed^{\D_i}$ & 1/6 & 2 & 3 \\
$ S^{\U_i}_R$ & 0 & 1 & $-\eed^{\U_i}$ & 2/3 & 1 & 3 \\
$ S^{\D_i}_R$ & 0 & 1 & $-\eed^{\D_i}$ & -1/3 & 1 & 3 \\
\hline
$\hat S^{\N_i}_L$ & 0 & 1 & $-\eed^{\N_i}$ & -1/2 & 2 & 1 \\
$\hat S^{\E_i}_L$ & 0 & 1 &  $-\eed^{\E_i}$ & -1/2 & 2 & 1 \\
$ S^{\N_i}_R$ & 0 & 1 &  $-\eed^{\N_i}$ & 0 & 1 & 1 \\
$ S^{\E_i}_R$ & 0 & 1 &  $-\eed^{\E_i}$ & -1 & 1 & 1 \\
		\rowcolor[gray]{.85}
        \multicolumn{7}{c}{\emph{Dark sector:}} \\
$Q^{\U_i}$  & 1/2 & 1 & $\eed^{\U_i}$ & 0 & 1 & 1 \\
$Q^{\D_i}$  & 1/2 & 1 & $\eed^{\D_i}$ & 0 & 1 & 1 \\
$Q^{\E_i}$  & 1/2 & 1 & $\eed^{\E_i}$ & 0 & 1 & 1 \\
$Q^{\N_i}$  & 1/2 & 1 & $\eed^{\N_i}$ & 0 & 1 & 1 \\
		\bottomrule
	\end{tabular}
  \caption{Particle content of the model and gauge assignments. The index $i=1,2,3$ runs over the SM generations. The electric charge of each field is given by $Q = I_3 + Y$, where $Y$ is the hypercharge and $I_3$ is the eigenvalue of the third weak isospin generator. }
	\label{tab:particles}
\end{table}

Concretely, we enlarge the SM gauge group by a $\mathbb{Z}_2$ discrete symmetry under which the fields transform as specified in Table~\ref{tab:particles}. The full Lagrangian is then given by  
\bea
    {\cal L}&=&{\cal L}_{\rm SM}(Y_f=0) + {\cal L}_{\rm MS}(q)+
    {\cal L}_{\rm MS}(L)+ {\cal L}_{\rm Dark}-V_{H_S}-V_{\rm MS}\,,
    \label{lagrangian}
\eea
where ${\cal L}_{\rm SM}(Y_f=0)$ represent the SM Lagrangian without the usual Yukawa interactions, which necessarily vanish for the considered $\mathbb{Z}_2$ assignments. The remaining terms host portal interactions and the dark sector fields, in particular ${\cal L}_{\rm MS}(q)$ and ${\cal L}_{\rm MS}(L)$ contain the Lagrangian for the messenger sector, including the portal interactions with quarks and leptons, respectively. The term ${\cal L}_{\rm Dark}$, instead, contains a set of massive Dirac fermions, the dark fermions, singlet under the SM gauge group but charged under a vectorial $U(1)_D$ dark gauge theory.
Next, $V(H_S)$ collects the terms of the scalar potential that involve the scalar field $H_S$, responsible for the spontaneous breaking of the new $Z_2$ symmetry. Explicitly
\bea
V_{H_S}=\lambda_{H_S} \frac{H_S^4}{4}-\mu_{S}^2\frac{H_S^2}{2} +\frac{1}{2}\lamHHS H_S^2 \hat{H}^{\dag} \hat{H}\, ,
\label{VHS}
\eea
where $\hat{H}$ stands for the SM Higgs doublet. As both $\hat{H}$ and $H_S$ develop non-vanishing VEVs, the last term in Eq.~\eqref{VHS} results in a tree-level mass mixing between the two scalars. In fact, focusing for the moment on the two Higgs fields only, the minimization conditions for the corresponding scalar potential set
\begin{align}
  \mu^2_S &= v^2 \lamHHS + v_S^2  \lambda_{H_S} \\
  \mu_H^2 &= v^2 \lambda_H + v^2_S \lamHHS\, 
\end{align}
and the matrix of squared masses of the $CP$-even bosons thus is: 
\begin{equation}
  \operatorname{M}^2
  =
  \begin{pmatrix}
  2 \, v^2 \, \lambda_H & 2 v \, v_S \, \lamHHS \\ 2 v \, v_S \, \lamHHS & 2 \, v_S^2 \, \lambda_{H_S}
  \end{pmatrix}\,.   
\end{equation}
The mass eigenstates are determined upon a rotation of the original scalar fields by an angle of 
\begin{equation}
  \tan 2\theta = \frac{2\, v\, v_S \,\lamHHS}{v^2\, \lambda_H - v_S^2 \, \lambda_{H_S}}\,.
\end{equation} 
Because in the rest of the paper we consider scenarios where $v_S \gg v$, we expand the above relation in powers of $v/v_S$, obtaining at the first order that  
\begin{equation}
  \tan 2\theta \approx - 2 \frac{\lamHHS}{\lambda_{H_S}} \frac{v}{v_S}\,.
\end{equation} 
As we can see, the effects of mass mixing in the considered limit are suppressed by the large hierarchy between the EW and NP scales and thus can be safely neglected. Still, because the $\lamHHS$ coupling receives important radiative contributions form the interactions contained in the ${\cal L}_{\rm MS}(q)$ and ${\cal L}_{\rm MS}(L)$ terms of Eq.~(\ref{lagrangian}), we retain the full RG evolution of this coupling in our analyses.

Finally, the last contribution in Eq.~\eqref{lagrangian},  $V_{\rm MS}$, contains the full scalar potential for the messenger fields and is separately discussed in the following.

The portal interactions in ${\cal L}_{MS}(q)$, responsible for the radiative generation of Yukawa couplings, are shaped by the SM quantum numbers and transmit the chiral symmetry breaking sourced by the dark fermion masses to quarks and leptons~\cite{Gabrielli:2013jka}.

In more detail, for the quark sector we have
\bea
    {\cal L}_{MS}(q)&=&{\cal L}^{0}_{MS}(q)+{\cal L}^I_{MS}(q)\, ,
    \eea
where ${\cal L}^{0}_{MS}(q)$ contains the kinetic terms, mass parameters and gauge interactions of the messenger fields and ${\cal L}^I_{MS}(q)$ specifies the portal interactions with the SM quarks:
\bea
{\cal L}^I_{MS}(q) &=&
\gUL \sum_{i=1}^{3}\left[\bar{q}^i_L Q_R^{\U_i}\right] \hat{S}^{\U_i}_{L} +
\gDL \sum_{i=1}^{3}\left[\bar{q}^i_L Q_R^{\D_i}\right] \hat{S}^{\D_i}_{L}
\nonumber\\
&+&
\gUR \sum_{i=1}^{3}\left[\bar{U}^i_R Q_L^{\U_i}\right] S^{\U_i}_{R} +
\gDR\sum_{i=1}^{3} \left[\bar{D}^i_R Q_L^{\D_i}\right] S^{\D_i}_{R} 
\nonumber\\
&+& \;
\lamSU \sum_{i=1}^{3}\tilde{H}^{\dag} \hat{S}^{\U_i}_L S^{\U_i\dag}_R H_S
+ \lamSD \sum_{i=1}^{3} \hat{H}^{\dag} \hat{S}^{\D_i}_L S^{\D_i\dag}_R H_S
\,+\, H.c. \, .
\label{LagMSq}
\eea 
In the equation above, we have left all the color and $SU(2)_{L}$ contractions understood and we have indicated the chiral projections of the dark fermion fields $Q^{\D_i}$ and $Q^{\U_i}$ with a subscript $L,R$. All sums run over $i=1,2,3$, corresponding to the SM fermion generations. The $SU(2)_{L}$ doublets $q^i_{L}=\left( U^i_{L}\, D^i_{L} \right)^T$ represent the SM up ($U$) and down ($D$) quark fields,
$\hat{S}^{\U_i,\D_i}_{L}=\left(S^{\U_i,\D_i}_{L_1} \, S^{\U_i,\D_i}_{L_2} \right)^T$, and $\hat H =\left(H^{+} \, H^{0} \right)^T$
is the SM Higgs doublet. As usual, $H^0 = (v + H) / \sqrt 2$ and the conjugate doublet is $\tilde{H}=i\sigma_2 \hat{H}^{\star}$. The fields carrying an $R$ subscript are $SU(2)_{L}$ singlets, including the complex scalar fields $S^{\U_i,\D_i}_{R}$. The constants $g^{\scriptscriptstyle{U,D}}_{\scriptscriptstyle{L}}$ and $g^{\scriptscriptstyle{U,D}}_{\scriptscriptstyle{R}}$ in Eq.~\eqref{LagMSq} are flavor-universal parameters that we require to lie in the perturbative regime throughout the following analysis.

The Lagrangian ${\cal L}_{MS}(L)$ that connects leptons to the corresponding dark fermions possesses a similar structure:
\bea
{\cal L}^I_{MS}(L) &=&
\gNL   \sum_{i=1}^{3}\left[\bar{L}^i_L Q_R^{\N_i}\right] \hat{S}^{\N_i}_{L} +
\gEL \sum_{i=1}^{3}\left[\bar{L}^i_L Q_R^{\LL_i}\right] \hat{S}^{\E_i}_{L}
\nonumber\\
&+&
\gNR \sum_{i=1}^{3}\left[\bar{\nu}^i_R Q_L^{\N_i}\right] S^{\N_i}_{R} +
\gER \sum_{i=1}^{3} \left[\bar{E}^i_R Q_L^{\LL_i}\right] S^{\E_i}_{R}
\nonumber\\
&+& \;
\lamSN \sum_{i=1}^{3}\tilde{H}^{\dag} \hat{S}^{\N_i}_L S^{\N_i\dag}_R H_S
+ \lamSL \sum_{i=1}^{3} \hat{H}^{\dag} \hat{S}^{\LL_i}_L S^{\E_i\dag}_R H_S
\,+\, H.c. \, .
\label{LagMSl}
\eea 
Here $L^i_{L}=\left(\nu^i_{L}\,E^i_{L} \right)^T$ represent the SM lepton doublets, with $E^i$ and $\nu_i$ being the charged lepton and neutrino fields, respectively. Similar to the case of quarks, we have $\hat{S}^{\N_i,\E_i}_{L}=\left(S^{\N_i,\E_i}_{L_1}\, S^{\N_i,\E_i}_{L_2} \right)^T$. The particle content of the SM has also been extended with right-handed neutrinos so that these particles can acquire mass in the same way as the remaining SM fermions, that is via effective Yukawa couplings. The framework is therefore compatible with the presence of three light \emph{Dirac} neutrinos.

The Lagrangian terms in Eqs.~(\ref{LagMSq}) and~(\ref{LagMSl}) contain the minimal set of interactions needed to produce the SM Yukawa couplings radiatively.
Due to the fact that the dark fermions $Q^{\U_i,\D_i,\N_i,\E_i}$ are SM gauge singlets, the quantum numbers of the messenger scalar fields necessarily coincide with those of squarks and slepton of supersymmetric models.

According to the proposals in Refs.~\cite{Gabrielli:2013jka,Gabrielli:2016vbb}, the assumption of flavor (generation) universality for the $g^{\U, \D,\E, \N}_{\LL,\RR}$, $\lambda_S^{\U, \D,\E, \N}$ couplings appearing in Eqs.~(\ref{LagMSq}) and (\ref{LagMSl}) attributes a potential flavor dependence of the Yukawa couplings in the quark or lepton sector solely to the involved dark-fermion masses. In fact, flavour universality is also preserved by the one-loop corrections to the mentioned couplings since only the dark-fermion masses break the universality. As a result, the spread of SM Yukawa couplings is directly related to the dark fermion mass spectrum, regardless of its origin. For instance, the use of non-perturbative dynamics in the dark sector easily allows exponentially spread dark fermions masses. The mechanism can therefore generate hierarchical SM Yukawa couplings that naturally fit the observed values, thereby solving the SM flavor puzzle~\cite{Gabrielli:2013jka,Gabrielli:2016vbb,Gabrielli:2019sjg}. Since in the present paper we are mainly concerned with the stability of the EW vacuum, which is not sensitive to the details of the flavour structure, we restrict ourselves to the generation of flavor-diagonal Yukawa couplings.

After the spontaneous symmetry breaking operated by the SM Higgs doublet and by the $H_S$ field, the last term in Eqs.~(\ref{LagMSq}) and (\ref{LagMSl}) result in the following trilinear couplings  
\bea
    {\cal L}_{3} &\supset& \lamSU v \sum_{i=1}^{3}\hat{S}^{\U_i}_L S^{\U_i\dag}_R H_S
+  \lamSD v\sum_{i=1}^{3} \hat{S}^{\D_i}_L S^{\D_i\dag}_R H_S
\,+
\lamSU v_S \sum_{i=1}^{3}\tilde{H}^{\dag} \hat{S}^{\U_i}_L S^{\U_i\dag}_R 
+ \lamSD v_S\sum_{i=1}^{3} \hat{H}^{\dag} \hat{S}^{\D_i}_L S^{\D_i\dag}_R 
\nonumber \\
&+&
\lamSN v \sum_{i=1}^{3}\hat{S}^{\N_i}_L S^{\N_i\dag}_R H_S
+ \lamSL v \sum_{i=1}^{3} \hat{S}^{\E_i}_L S^{\E_i\dag}_R H_S
\,+
\lamSN v_S \sum_{i=1}^{3}\tilde{H}^{\dag} \hat{S}^{\N_i}_L S^{\N_i\dag}_R 
+ \lamSL v_S \sum_{i=1}^{3} \hat{H}^{\dag} \hat{S}^{\E_i}_L S^{\E_i\dag}_R  
\nonumber \\
&+& H.c. \,  ,
\label{Vtrilinear}
\eea
where $v_S$ is the VEV of the $H_S$ field. As we show in the next section, these trilinear couplings are crucial for the radiative generation of the SM Yukawa couplings.

To conclude the section, we report below the most general expression for the minimal form of quartic potential $V_{\rm MS}$ of messenger scalar fields allowed by the symmetries of the theory. The expression takes into account the following aspects:
\begin{itemize}
\item[i)] the hypothesis of flavor universality of the $g^i_{\scriptscriptstyle{L,R}}$, $\lambda^i_S$ couplings, $i=\scriptstyle{U,D,E,N}$, in Eqs.~(\ref{LagMSq}) and (\ref{LagMSl}). 
\item [ii)]
  the fact that the messengers and dark-fermions are both charged under $U(1)_D$ gauge interactions, with different $U(1)_D$ charges.
\end{itemize}
  The above conditions result in a radiatively generated potential given by
\bea
    {V}_{\rm MS}&=&
    \lamULL \sum_{i=1}^3 \left[\left(\hat S_{L}^{\U_i}\right)^{\dag} \hat S^{\U_i}_{L}\right]^2 + \lamDLL \sum_{i=1}^3 \left[\left(\hat S_{L}^{\D_i}\right)^{\dag} \hat S^{\D_i}_{L}\right]^2 \nonumber \\
    &+&  \lamURR \sum_{i=1}^3 \left[\left(S_{R}^{\U_i}\right)^{\dag} S^{\U_i}_{R}\right]^2 + \lamDRR \sum_{i=1}^3 \left[\left(S_{R}^{\D_i}\right)^{\dag} S^{\D_i}_{R}\right]^2\nonumber \\
&+&    \lamULR \sum_{i=1}^3 \left(\hat S_{L}^{\U_i}\right)^{\dag} \hat S^{\U_i}_{L}\left(S_{R}^{\U_i}\right)^{\dag} S^{\U_i}_{R}+
    \lamDLR \sum_{i=1}^3 \left(\hat S_{L}^{\D_i}\right)^{\dag} \hat S^{\D_i}_{L}\left(S_{R}^{\D_i}\right)^{\dag} S^{\D_i}_{R}\nonumber \\
    &+&  \lamNLL \sum_{i=1}^3 \left[\left(\hat S_{L}^{\N_i}\right)^{\dag} \hat S^{\N_i}_{L}\right]^2 + \lamELL \sum_{i=1}^3 \left[\left(\hat S_{L}^{\E_i}\right)^{\dag} \hat S^{\E_i}_{L}\right]^2 \nonumber \\
    &+&  \lamNRR \sum_{i=1}^3 \left[\left(S_{R}^{\N_i}\right)^{\dag} S^{\N_i}_{R}\right]^2 + \lamERR \sum_{i=1}^3 \left[\left(S_{R}^{\E_i}\right)^{\dag} S^{\E_i}_{R}\right]^2 \nonumber \\
    &+&    \lamNLR \sum_{i=1}^3 \left(\hat S_{L}^{\N_i}\right)^{\dag} \hat S^{\N_i}_{L}\left(S_{R}^{\N_i}\right)^{\dag} S^{\N_i}_{R}+
    \lamELR \sum_{i=1}^3 \left(\hat S_{L}^{\E_i}\right)^{\dag} \hat S^{\E_i}_{L}\left(S_{R}^{\E_i}\right)^{\dag} S^{\E_i}_{R}\, .
\label{VMS}
\eea

The number of independent parameters included in the expression above is the minimal compatible with the symmetries of the theory and flavor universality. The number of couplings could be further reduced only by assuming extra symmetries. For instance, the LR gauge symmetry considered in Ref.~\cite{Gabrielli:2016vbb} forces $\lambda^{\U, \D,\E, \N}_{\scriptscriptstyle{LL}}=\lambda^{\U, \D,\E, \N}_{\scriptscriptstyle{RR}}$  and  $g^{\U, \D,\E, \N}_{\scriptscriptstyle{L}}=g^{\U, \D,\E, \N}_{\scriptscriptstyle{R}}$ at the high energy scale where the LR symmetry is first broken. In our case, these relations are not preserved by radiative corrections because the $L-$ and $R$-type messenger fields have different $SU(2)_L\times U(1)_Y$ quantum numbers.

In the following analysis we also disregard all quartic couplings of the form $\hat{H}^{\dag} \hat{H} \hat S^{I \dag}_L \hat S_L^I$ or  $\hat{H}^{\dag} \hat{H} S^{I \dag}_R S_R^I$, involving  two messenger fields of the same type (I runs over the SM fields) and the SM Higgs doublet $\hat{H}$ or the singlet $H_S$.
Even if vanishing at a scale, these couplings are inevitably re-generated by radiative corrections already at the one-loop level. However, with the interactions of the Higgs boson and $H_S$ field included in Eq.~\eqref{LagMSl}, the $\beta$-functions of the couplings that multiply the  $\hat{H}^{\dag} \hat{H} S^{I \dag}_X S_X^I$ operators, $X=L,R$, receive a first contribution proportional to the square of the EW gauge couplings at the one-loop level.  Operators involving $H_S$, instead, begin to run only at higher orders. By setting these parameters to small and positive values at a given scale, the slow running then ensures that their effect on the RGEs of the model -- and in particular on the evolution of the Higgs quartic coupling -- is always negligible. Consequently, we expect that a more careful assessment of the RGE evolution in the present model would only marginally change the results of our vacuum stability analysis.

\section{Radiative generation of Yukawa couplings}
\label{sec:3}

We begin our investigation by identifying the couplings and mass scales relevant to the problem of vacuum stability, using the simplified framework introduced in the previous section as a benchmark. 

\begin{figure}[h]
  \centering
    \includegraphics[width=.7\textwidth]{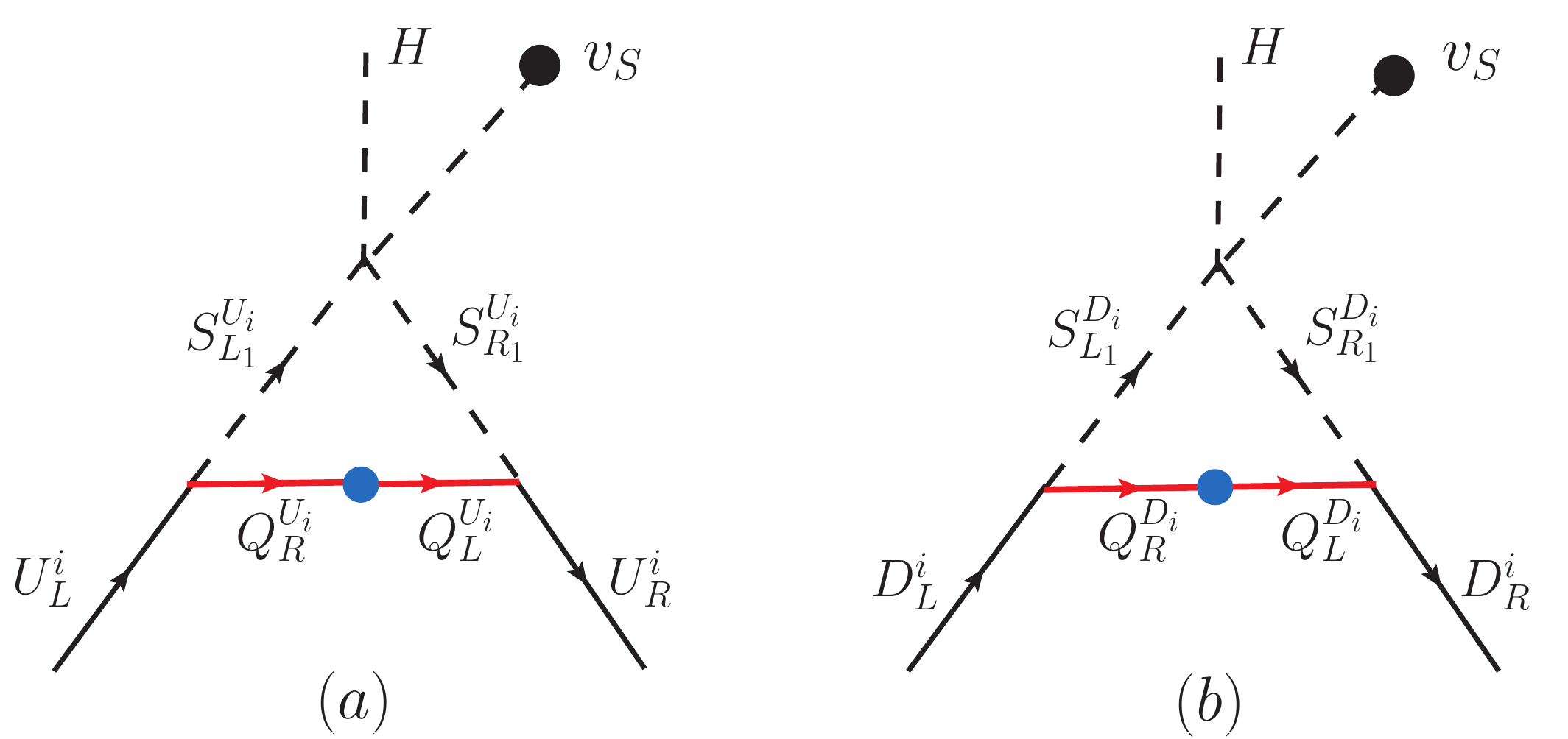}
  \caption{\it The diagrams responsible for the radiative generation of the Yukawa couplings of up-type quarks (a) and down-type quarks (b). The black circle on the external scalar line implies that the field $H_S$ is set to its vacuum expectation value $v_S$. The blue dot on the dark fermion line (in red), instead, represents a mass insertion.}
  \label{fig1}
\end{figure}

The SM Yukawa couplings arise at the one-loop level through diagrams analogous to that of Fig.~\ref{fig1}, which explicitly show the case of quarks. 

The expressions for the resulting SM Yukawa couplings can be obtained by matching the fundamental 5-dimensional amplitudes to the local SM Yukawa operators. In particular, for the quark sector we have~\cite{Gabrielli:2013jka}
\bea
Y^q_i &=& \frac{g^q_{\LL} g^q_{\RR}}{16\pi^2} \left(\frac{M_{Q_i} \Lambda_S}{m^2}\right)\, f_1(x_i,\xi)
\label{yukawa}
\eea
where the index $i$ stands for the quark flavor (for both up and down types) and  $f_1(x,\xi)$ is a loop function given by
\bea
f_1(x,\xi)&=&\frac{1}{2}\left[
C_0\left(\frac{x}{1-\xi}\right)\frac{1}{1-\xi}+C_0\left(\frac{x}{1+\xi}\right)\frac{1}{1+\xi}\right]\,,
\label{f1}
\eea 
with
\bea
C_0(x)=\frac{1-x\left(1-\log{x}\right)}{(1-x)^2}\, .
\label{C0}
\eea
In the above equations we have used $x_i= M_{Q_i}^2/m^2$, where $m^2$ is the average mass of the colored scalar messengers and $M_{Q_i}$ is the mass of the dark fermion associated to the quark $q^i$. We have also defined
\bea
\Lambda_S &=& \lambda_S v_S\, ,
\nonumber \\
\xi&=&\frac{\Lambda_S v}{m^2}\, ,
\label{LambdaS-xi}
\eea
which indicate the scale of new physics and the strength of the mixing in the colored messenger mass sector, respectively.

The request that messenger fields be unstable, or analogously that the dark fermions be stable particles, forces the latter to be lighter than the former. The condition translates into the following bound
\bea
M^2_{Q^i} < m^2(1-\xi)\, ,
\eea
where we have neglected the contributions of quark masses, subdominant with respect to those of dark-fermions.

In analogy to the above results, the SM Yukawa couplings of leptons are
\bea
Y^{\E,\N}_i &=& \frac{g^{\E,\N}_{\LL} g^{\E,\N}_{\RR} }{16\pi^2} \left(\frac{M_{Q_i} \Lambda_S}{\bar{m}^2}\right)\, f_1(x,\bar{\xi})
\label{yukawaL}
\eea
where now $M_{Q_i}$ is the mass of the dark fermion associated to the lepton of generation $i$ and $\bar{m}$ and $\bar{\xi}$ are the common mass and the mixing angle in the leptonic sector of messengers, respectively. 

Without loss of generality, henceforth we set the masses of all the messenger fields to a common scale, imposing $m=\bar{m}$, and likewise require that $\xi=\bar{\xi}$. In fact, these simplifications do not preclude the framework from reproducing arbitrary Yukawa hierarchies, which can be matched by considering suitable dark fermion mass spectra and rescaling of the $g^q_{\LL,\RR}$, $g_{\LL,\RR}^{\E}$ and $g_{\LL,\RR}^{\N}$ couplings. Notice that since avoiding tachyons and color- or charge-breaking minima in the messengers sector requires $\xi<1$, we can bound the scale of new physics through
  \bea
  \Lambda_S < \frac{m^2}{v}\,.
  \eea

In order to further reduce the parameter space, we require that $g^{q}_{\LL}=g_{\RR}^{q}\equiv g_{\LR}^q$, $g^{\N,\E}_{\LL}=g_{\RR}^{\N,\E}\equiv g_{\LR}^{\LL}$ at the energy scale $\mu_{\rm mes}\sim m$ of the order of the common messenger mass. Since the messengers are the heaviest fields running inside the relevant loop contributions, we match the fundamental amplitude of Fig.~\ref{fig1} to the SM Yukawa operator at the same scale $\mu_{\rm mes}$. 

A closer inspection of Eq.~(\ref{yukawa}) reveals that the radiative suppression factor is to be compensated by large couplings $g^q_{\LL,\RR}\simeq {\cal O}(1)$ to reproduce the observed value of $Y_t\sim 1$. Alternatively, it is possible to take $\Lambda_S/m \gg 1$ (but $\Lambda_S/m< m/v$ to avoid tachyons) and match the value of top Yukawa coupling for perturbative values of $g^q_{\LL,\RR}$. Indeed, large values of the trilinear coupling $\Lambda_S$ are allowed at high energy because the associated operator has dimension three. However, they can break the perturbative unitarity of the $S$ matrix at low energies~\cite{DiLuzio:2016sur}. In particular, $\Lambda_S$ appears in the interaction vertex $\Lambda_S H S^{\dag}_i S_i$  between generic messenger fields denoted by the index $i$. Large values of the parameter can therefore cause the elastic scatterings cross section for the $S_i S_i\to S_i S_i$, mediated by the SM Higgs boson, to grow beyond the unitarity limit. This might signal 
the formation of bound states of messenger fields, which would allow to recover the unitarity of the theory in a non-perturbative way.

The issue of perturbative unitarity in the radiative generation of $Y_t$ becomes evident once the value of the dark fermion mass associated to the top quark is chosen so as to maximize the radiative contributions to $Y_t$. Setting the parameter to the same order of the lightest messenger mass, $M_{Q_t}\sim m$, implies $x_t= 1-\xi$ and gives
\bea
Y_t &=& \frac{(g^q_{\LR})^2}{16\pi^2}\left(\frac{m}{v}\right)F_Y(\xi)\, .
\label{yukawatop}
\eea
The function $F_Y(\xi)$ is 
\bea
F_Y(\xi)&=&\frac{2\xi+(1-\xi)^2\log{\left(\frac{1-\xi}{1+\xi}\right)}}{8\xi\sqrt{1-\xi}}\, ,
\label{Ft}
\eea
and admits the limit $F_Y(\xi)\simeq \frac{\xi}{2}+{\cal O}(\xi^2)$ in the small mixing case, $\xi\ll 1$. We can then express the Yukawa coupling as a function of the $\Lambda_S$ scale
\bea
Y_t&\sim &\frac{(g^q_{\LR})^2}{32 \pi^2} \frac{\Lambda_S}{m} + {\cal O}(\xi^2)\,,
\label{Ytopapprox}
\eea
and the value of $Y_t$ can be matched by adjusting the product $(g^q_{\LR})^2 \Lambda_S$. For the purpose of assessing the vacuum stability, we can then use the relation in Eq.~(\ref{yukawatop}) to infer the initial condition for $g^q_{\LR}$, given at the messengers scale as a function of $m$ and $\xi$. Equivalently, in the small mixing regime, we can use Eq.~(\ref{Ytopapprox}) to determine the value of $\Lambda_S/m$ as a function of $g_{\LR}^q$.

The same strategy also allows to determine the $g^{{\LL}}_{\LR}$ couplings, although perturbative unitarity can be easily respected in the leptonic sector  because $Y_{\tau} \ll 1$. Requiring that the mass of the dark-fermion associated to the tau lepton be of the same order of the lightest messenger mass, we then have
\bea
Y_{\tau} &=& \frac{(g^{\LL}_{\LR})^2}{32\pi^2}\left(\frac{m}{v}\right)
F_Y(\xi)\, , 
\label{yukawatau}
\eea
where, as anticipated, we have used the same average messenger mass and mixing as in the quark sector. For the measured values of SM Yukawa couplings, we do not expect the simplification to induce qualitative changes in the RG evolution of the SM Higgs boson quartic coupling because the hierarchy between quark and lepton messengers spans at most two orders of magnitude.

In the following, after detailing the relevant $\beta$-functions, we analyze the EW vacuum stability in two complementary scenarios delineated by the above considerations:
\begin{itemize}
  \item[I)]
We consider perturbative values of the couplings $g^{{\LL},q}_{\LR} \lesssim 1$ and a set of values for the common messenger mass $m$. The scale $\Lambda_S$ is then adjusted so as to reproduce the observed value of $Y_{t}$ through Eq.~(\ref{yukawatop}) and we extend the vacuum stability analysis up to the Planck scale $\mu \sim M_{\rm Pl}$. Due to the large ratio $\Lambda_S/m \gg 4\pi$, we assume that the non-perturbative phenomena needed to recover unitarity at low energy in messenger sector do not affect the running of $\lambda_H$ at the large scales relevant for the vacuum stability. This is justified by the fact that operators of dimension 3, potentially responsible for breaking unitarity at low energy, are super-renormalizable in the UV. We also speculate on a possible UV completion which allows to have a large ratio $\Lambda_S/m \gg 1$ compatible with perturbative couplings $g^q_{\LR} \ll 1$ at low energy.

\item[II)]
We use $\Lambda_S/m \lsim 4\pi$, within the limit of perturbative unitarity. The initial values of the couplings $g^{q}_{\LR}$ and $g^{{\LL}}_{\LR}$ are then extracted from $Y_t$ and $Y_{\tau}$ in Eq.~(\ref{yukawatop}). Because  $g^{q}_{\LR}$ is necessarily borderline with the perturbative limit, we analyze the running of $\lambda_H$ only up to the scale where the first Landau pole is reached.
\end{itemize}

In both scenarios, we regard the common messenger mass, $m$, and the trilinear Higgs-messengers coupling, $\Lambda_S$, as input parameters. For the sake of simplicity, we set the quartic coupling $\lambda_{H_S}$ in a way that the mass of the $H_S$ field matches the common messenger mass scale.  Beside the quantities that directly determine the Higgs boson quartic coupling, we track the RG evolution of the remaining couplings to ensure the absence of color breaking\footnote{The emergence of color breaking minima is prevented by requiring that quartic couplings involving messenger fields remain positive at all scales. This also prevents the appearance of mass mixing terms involving messenger fields and the SM Higgs boson, which would be otherwise generated through the $\hat{H}^{\dag} \hat{H} S^{I \dag}_X S_X^I$ operators, $X=L,R$, neglected in this analysis.} and assess their perturbativity.

\subsection{Further phenomenological implications}
We conclude the Section with a brief review of the phenomenological implications of the framework, based on the works of Refs.~\cite{Biswas:2017lyg,Biswas:2016jsh,Biswas:2015sha,Gabrielli:2014oya,Fabbrichesi:2017zsc,Gabrielli:2016cut,Barducci:2018rlx,Fabbrichesi:2017vma, Gabrielli:2019sjg}.

In this scenario, the generation of the SM Yukawa interactions requires the existence of heavy scalar messenger fields and light dark-fermions, both charged under an unbroken $U(1)_D$ gauge interaction in the dark sector. Importantly, the model then clearly allows for the direct production of a pair of colored scalar messenger fields at collider experiments, via gluon-gluon fusion or quark-antiquark annihilation (the latter proceeding through the exchange of dark fermions in the t- or u-channel). Each messenger field eventually decays into the corresponding quark and dark-fermion, resulting thereby in a jet accompanied by missing energy. This signature is quite similar to the squark production of supersymmetric models with a stable neutralino, which plays here the role of a dark fermion. Although a dedicated collider analysis is still missing, we expect that the sensitivity of the experiment to the cross sections will be reduced by the mass of the messenger fields with respect to the corresponding supersymmetric case. In particular, the LHC can only probe the direct production of messenger fields with masses up to a few TeV.

The framework also foresees the existence of a light sector containing the massless dark photon, $\bar{\gamma}$, associated to the dark $U(1)_D$ gauge symmetry. This $U(1)_D$ guarantees the stability of dark fermions, required by DM phenomenology, and protects the theory from inducing large tree-level flavour-changing neutral current (FCNC) transitions. Although the dark photon does not couple to  ordinary matter at the tree-level, effective couplings are generated by higher dimensional operators involving quarks and leptons in the loop. Then, another distinguishing feature of this scenario is the predicted decay of the Higgs boson into photon and dark photon $H\to \gamma \bar{\gamma}$~\cite{Gabrielli:2014oya,Biswas:2016jsh}, which gives rise to a monochromatic photon plus (neutrino-like) missing energy signature at the LHC~\cite{Gabrielli:2014oya,Biswas:2016jsh} or future $e^+e^-$ colliders~\cite{Biswas:2015sha}. This process is induced at the one-loop level by the exchange of messenger fields in the loop. For the non decoupling properties of the Higgs boson, we expect that sizeable ratios of the percent level could be achieved even for very large masses of the messenger fields. In regard of this, the ATLAS~\cite{ATLAS:2021pdg,ATLAS:2021gly} and CMS~\cite{CMS:2020krr,CMS:2020hqe,CMS:2019ajt}
collaborations have recently begun their investigation of this signature producing quite stringent upper bounds on the process.
  
Another signature of the model are the FCNC processes induced by the decay of a SM fermion ($f$) into a lighter one ($f^{\prime}$) of same charge plus a dark-photon, $f\to f^{\prime} +\bar{\gamma}$~\cite{Gabrielli:2016cut}, active in both the quark and lepton sectors. In particular, implications for the dark photon production via the rare charged Kaon decay $K^+\to \pi^+ \pi^0+\bar{\gamma}$ (induced by $s\to d \bar{\gamma}$ transitions) have been analyzed in Ref.~\cite{Fabbrichesi:2017vma}. The large branching ratios expected for this decay could be of interest for experiments dedicated to rare $K^+$ decays like the NA62 at CERN.
 
Finally, the production of light dark-fermions in invisible decays of neutral hadrons has been investigated in Ref.~\cite{Barducci:2018rlx}, finding that the expected  branching ratios of the $K_L$ and $B^0$ mesons are comparable to the current experimental limits.

\section{RGEs for the full model}
\label{sec:4}

We present here the $\beta$-functions of the parameters that we track in our analysis of vacuum stability. Due to the approximations adopted and the precision used in the computation of the effective Yukawa couplings, it is sufficient to compute the corresponding RGEs at the 1-loop order.

In studying our benchmark model, we have assumed a common $U(1)_D$ charge for all the dark fermions and mediator fields expecting that the generalization to non-universal charges will not change the conclusion of the analysis. This assumption also guarantees the flavor (family) universality of the 1-loop RGEs.

The convention we use for the $\beta$-function is: 
\begin{equation*}
\beta\left(X\right) \equiv \mu \frac{d X}{d \mu}\equiv\frac{1}{\left(4 \pi\right)^{2}}\beta^{(1)}(X)\,.
\end{equation*}

In our analysis, we run the SM RGEs from the top quark mass scale to the matching scale $\mu=\mu_{\rm mes} \simeq m$, where the parameters of our benchmark model are initialized\footnote{Dark Matter phenomenology forces the dark fermions to be lighter than the messengers. Therefore, the scale $m$ corresponds to the largest scale associated to the degrees of freedom that circulate in the loop diagrams responsible for the Yukawa couplings generation.}. 
We then continue the RG evolution of the quantities under investigation by using the $\beta$-functions obtained for the benchmark model, up to a scale $\mu = \LambdaUV$ corresponding to the UV cutoff of the theory. As for this, in absence of a UV completion for gravitational interactions, it is customary~\cite{Gabrielli:2013hma} to assume as a UV cutoff the lowest  between the Planck mass, $M_{\rm Pl}$, and the  scale $M_{\rm LP}$ at which the first Landau pole appears in the evolution of a coupling:
\bea
\LambdaUV={\rm min}\left[M_{\rm Pl},M_{\rm LP}\right]\, .
\label{cutoff}
\eea
In particular, in the second scenario we consider, Landau poles might appear in the RG flow of $g_{\LL,\RR}^{q}(\mu)$ well below the Planck scale, $M_{\rm LP} < M_{\rm Pl}$, due to the large initial values of these couplings imposed by the matching with the SM top Yukawa coupling. The above criterion was introduced in Ref.~\cite{Gabrielli:2013hma} to investigate the stability of the SM vacuum under the assumption that quantum gravity does not introduce additional particle threshold above the Planck scale, as expected for instance in asymptotic safety scenarios~\cite{Dubovsky:2013ira}.

For the sake of convenience, we also report the SM 1-loop $\beta$-function for the Higgs boson quartic coupling used for the RG evolution of the parameter in the range  $m_{t}< \mu < \mu_{\rm mes}$, with $\mu_{\rm mes} \sim {\cal O}(m)$:
\bea
\label{eq:smlh}
\beta^{(1)}(\lambda_{H}) &=&
24 \lambda_{H}^{2}
-6 Y_t^4 +12\lambda_H Y_t^2
- \frac{9}{5} g^{\prime 2} \lambda_{H}
- 9 g^{2} \lambda_{H}
+ \frac{27}{200} g^{\prime 4}
+ \frac{9}{20} g^{2} g^{\prime 2}
+ \frac{9}{8} g^{4}\, .~~~~~
\eea

\subsection{Quartic couplings}
We provide below the 1-loop $\beta$-functions for the quartic couplings
of the model as defined in Section~\ref{sec:3},  valid for
$\mu_{\rm mes} < \mu < \LambdaUV$, with the scale $\LambdaUV$ as defined in Eq.~(\ref{cutoff}):

\bea
\label{eq:fmh}
\beta^{(1)}(\lambda_{H}) &=&
24 \lambda_{H}^{2}
+ \frac{1}{2} \lamHHS^{2}
- \frac{9}{5} g^{\prime 2} \lambda_{H}
- 9 g^{2} \lambda_{H}
+ \frac{27}{200} g^{\prime 4}
+ \frac{9}{20} g^{2} g^{\prime 2}
+ \frac{9}{8} g^{4}\, ,
\eea

\bea
\beta^{(1)}(\lambda_{H_S}) &=&
18 \lambda_{H_S}^{2}
+2\lamHHS^{2}\, ,
\eea

\bea
\beta^{(1)}(\lamHHS) &=&
4N_F \Big( 3 (\lamSU)^2+ 3 (\lamSD)^2+ (\lamSL)^2 + (\lamSN)^2\Big) 
+  \lamHHS\Big(
12\lambda_{H} 
+ 6 \lambda_{H_S}
+ 4 \lamHHS\nonumber \\
&-&  \frac{9}{10} g^{\prime 2}
-  \frac{9}{2} g^{2} \Big)\, ,
\eea

\bea
\beta^{(1)}(\lamSq) &=&\lamSq\left(
2 \lambda_{HH_s}
+ 2 \lamLRq
-  C^{q}_{\SSS} g^{\prime 2} 
-  \frac{9}{2} g^{2} 
- 8  g_{3}^{2} 
- 6 \gD^{2} 
+ \left|\gQL\right|^2
+ \left|\gQR\right|^2\right)\, ,
\eea

\bea
\beta^{(1)}(\lamSl) &=&\lamSl\left(
2 \lambda_{HH_s}
+ 2 \lamLRl
-  C^{\LL}_{\SSS} g^{\prime 2} 
-  \frac{9}{2} g^{2} 
- 6 \gD^{2} 
+ \left|\gLL\right|^2
+ \left|\gLR\right|^2\right)\, ,
\eea

\bea
\beta^{(1)}(\lamLLQ) &=&
\lamLLQ\left(
40 \lamLLQ
-\frac{1}{5} g^{\prime 2} 
- 9 g^{2} 
- 16 g_{3}^{2} 
- 12 \gD^{2} \right)
+ 3 (\lamLRq)^{2}
\\\nonumber
&+&  \frac{1}{600} g^{\prime 4}
+ \frac{1}{20} g^{2} g^{\prime 2}
+ \frac{9}{8} g^{4}
+ g^{2} g_{3}^{2}
+ 3 g^{2} \gD^{2}
+ \frac{1}{15} g_{3}^{2} g^{\prime 2}
+ \frac{13}{6} g_{3}^{4}
\\\nonumber
&+&  4 g_{3}^{2} \gD^{2}
+ \frac{1}{5} \gD^{2} g^{\prime 2}
+ 6 \gD^{4}
+ 4 \lamLLQ \left|\gQL\right|^{2}
- 2 \left|\gQL\right|^{4}\, ,
\eea

\bea
\beta^{(1)}(\lamLLL) &=&
\lamLLL\left(
24 \lamLLL
- \frac{9}{5} g^{\prime 2} 
- 9 g^{2} 
- 12 \gD^{2} \right)
+ (\lamLRl)^{2}
+ \frac{27}{200} g^{\prime 4}
+ \frac{9}{20} g^{2} g^{\prime 2}
+ \frac{9}{8} g^{4}
\\
\nonumber
&+& 3 g^{2} \gD^{2}
+ \frac{9}{5} \gD^{2} g^{\prime 2}
+ 6 \gD^{4}
+ 4 \lamLLL \left|\gLL\right|^{2}
- 2 \left|\gLL\right|^{4}\, ,
\eea

\bea
\beta^{(1)}(\lamRRQ) &=&
\lamRRQ\Big(
28 \lamRRQ
-  (C_{\Y}^{q})^2\frac{4}{5} g^{\prime 2} 
- 16 g_{3}^{2} 
- 12 \gD^{2} \Big)
+ 6 (\lamLRq)^2
+ \frac{2}{75}(C_{\Y}^{q})^4 g^{\prime 4}
\\
\nonumber
&+& \frac{4}{15} (C_{\Y}^{q})^2 g_{3}^{2} g^{\prime 2}
+ \frac{13}{6} g_{3}^{4}
+ 4 g_{3}^{2} \gD^{2}
+ \frac{4}{5} (C_{\Y}^{q})^2 \gD^{2} g^{\prime 2}
+ 6 \gD^{4}
+ 4 \lamRRQ \left|\gQR\right|^{2}
- 2 \left|\gQR\right|^{4}\, ,
\eea

\bea
\beta^{(1)}(\lamRRL) &=&
\lamRRL\Big(
20 \lamRRL
- \frac{36}{5}  (C_{\Y}^{\LL})^2 g^{\prime 2} 
- 12 \gD^{2} \Big)
+ \frac{54}{25} (C_{\Y}^{\LL})^4 g^{\prime 4}
+ 2(\lamLRl)^2
\\
\nonumber
&+& \frac{36}{5} (C_{\Y}^{\LL})^2 \gD^{2} g^{\prime 2}
+ 6 \gD^{4}
+ 4 \lamRRL \left|\gLR\right|^{2}
- 2 \left|\gLR\right|^{4}\, ,
\eea

\bea
\beta^{(1)}(\lamLRq) &=&
2 (\lamSq)^2
+\lamLRq\Big(
28 \lamLLQ 
+ 16 \lamRRQ
+ 4 \lamLRq
-  C^{q}_{\LR} g^{\prime 2} 
-  \frac{9}{2} g^{2} 
- 16 g_{3}^{2} 
- 12 \gD^{2} \Big)
\\
\nonumber
&+& (C_{\Y}^{q})^2 \frac{1}{75} g^{\prime 4}
+ C_{\Y}^q \frac{4}{15} g_{3}^{2} g^{\prime 2}
+ \frac{13}{3} g_{3}^{4}
+ 8 g_{3}^{2} \gD^{2}
+ C_{\Y}^{q}\frac{4}{5} \gD^{2} g^{\prime 2}
+ 12 \gD^{4}
\\ \nonumber
&+& 2 \lamLRq \left|\gQL\right|^{2}
+ 2 \lamLRq \left|\gQR\right|^{2}\, ,
\eea

\bea
\beta^{(1)}(\lamLRl) &=&
2 (\lamSl)^2
+\lamLRl\Big(
12 \lamLLL 
+ 8 \lamRRL
+ 4 \lamLRl
-  C^{\LL}_{\LR} g^{\prime 2} 
-  \frac{9}{2} g^{2} 
- 12 \gD^{2} \Big)
\\
\nonumber
&+& (C_{\Y}^{\LL})^2 \frac{27}{25} g^{\prime 4}
+  C_{\Y}^{\LL} \frac{36}{5} \gD^{2} g^{\prime 2}
+ 12 \gD^{4}
+ 2 \lamLRl \left|\gLL\right|^{2}
+ 2 \lamLRl \left|\gLR\right|^{2}\, ,
\label{beta-quartic}
\eea

\noindent
where the superscript $q=\;\scriptstyle{U,D}$, $\scriptstyle{L=E,N}$, the couplings $g^{\prime}$, $g$, $g_3$, and $\gD$ correspond to the gauge groups $U(1)_Y$, $SU(2)_L$, $SU(3)_c$ and $U(1)_D$ respectively, and $N_F=3$ is the number of SM generations or families. The constants coefficients that differentiate between the $\beta$-functions are $C^{\U}_{\SSS}=13/10$, $C^{\D}_{\SSS}=7/10$, $C^{\E}_{\SSS}=27/10$, $C^{\N}_{\SSS}=9/10$, $C_{\Y}^{\U} = 2$, $C_{\Y}^{\D} = -1$,  $C_{\Y}^{\E} = 1$,  $C_{\Y}^{\N} = 0$, $C^{\U}_{\LR} = 17/10$,  $C_{\LR}^{\D} = 1/2$, $C^{\E}_{\LR} = 9/2$, $C^{\N}_{\LR}=9/10$.

Notice that the large negative contribution of the top quark Yukawa coupling to $\beta^{(1)}(\lambda_{H})$ (corresponding to $-6Y_t^2$ in Eq.~\eqref{eq:smlh}), which is the main cause of vacuum instability in the SM, vanishes above the messenger scale. The new contribution to the RGE of $\lambda_{H}$  is given by the positive term proportional to $\lamHHS^2$.

\subsection{Dark Yukawa couplings}
The $\beta$-function for the flavor universal couplings $g_{{\LL},{\RR}}^{\X}$ of the model, $\scriptstyle{X = U,D,E,N}$, defined in Section~\ref{sec:3} and valid for $\mu_{\rm mes} < \mu < \LambdaUV$, are:
\bea  
\beta^{(1)}(\gUL) &=&
\gUL\left(
\frac{9}{2} \left|\gUL\right|^{2}
+ \frac{1}{2} \left|\gDL\right|^{2}
-  \frac{1}{20} g^{\prime 2} 
-  \frac{9}{4} g^{2} 
- 4 g_{3}^{2} 
- 3 \gD^{2}
\right)
\eea

\bea
\beta^{(1)}(\gUR) &=&
\gUR\left(
3 \left|\gUR\right|^{2}
-  \frac{4}{5} g^{\prime 2} 
- 4 g_{3}^{2} 
- 3 \gD^{2}
\right)
\eea

\bea
\beta^{(1)}(\gDL) &=& \gDL\left(
\frac{9}{2} \left|\gDL\right|^{2}
+  \frac{1}{2}   \left|\gUL\right|^{2}
-  \frac{1}{20} g^{\prime 2} 
-  \frac{9}{4} g^{2} 
- 4 g_{3}^{2} 
- 3 \gD^{2} \right)
\eea

\bea
\beta^{(1)}(\gDR) &=&\gDR\left(
3 \left|\gDR\right|^{2}
-  \frac{1}{5} g^{\prime 2}
- 4 g_{3}^{2}
- 3 \gD^{2}
\right)
\eea

\bea
\beta^{(1)}(\gNL) &=&
\gNL\left(
\frac{5}{2}  \left|\gNL\right|^{2}
+ \frac{1}{2} \left|\gEL\right|^{2}
-  \frac{9}{20} g^{\prime 2} 
-  \frac{9}{4} g^{2} 
- 3 \gD^{2}
\right)
\eea

\bea
\beta^{(1)}(\gNR) &=&
\gNR\left(
2 \left|\gNR\right|^{2}
-  3g_D^{2} 
\right)
\eea

\bea
\beta^{(1)}(\gEL) &=& \gEL\left(
 \frac{1}{2} \left|\gNL\right|^{2}
+  \frac{5}{2} \left|\gEL\right|^{2}
-  \frac{9}{20} g^{\prime 2} 
-  \frac{9}{4} g^{2}  
- 3 \gD^{2} \right)
\eea

\bea
\beta^{(1)}(\gER) &=&\gER\left(
2 \left|\gER\right|^{2}
- \frac{9}{5} g^{\prime 2}
- 3 \gD^{2}
\right)\, .
\eea

\subsection{Scalar mass parameters}
Finally, we report the $\beta$-functions for the mass parameters of the considered scalar fields. 
\bea
\beta^{(1)}(\mu_S^2) &=&6 \lambda_{H_S}\mu_S^2
+ 4 \lamHHS \mu_H^2
\eea
\bea
\beta^{(1)}(\mu_H^2) &=&\mu_H^2\left(
-  \frac{9}{10} g^{\prime 2} 
-  \frac{9}{2} g^{2} 
+ \frac{\lamHHS \mu_S^{2}}{\mu_H^2}
+ 12 \lambda_{H} \right)
\eea
\bea
\beta^{(1)}(m^2_{S^q_L}) &=&  6 \lamLRq m^2_{S^q_R}
+ m^2_{S^q_L}\left(
-  \frac{1}{10} g^{\prime 2} 
-  \frac{9}{2} g^{2} 
- 8 g_{3}^{2} 
- 6 \gD^{2} 
+ 28 \lamLLQ
+ 2 \left|{\gQL}\right|^{2}\right)\, 
\eea
\bea
\beta^{(1)}(m^2_{S^L_L}) &=&  2 \lamLRl m^2_{S^L_R}  
+ m^2_{S^L_L}\left(
-  \frac{9}{10} g^{\prime 2} 
-  \frac{9}{2} g^{2} 
- 6 \gD^{2} 
+ 12 \lamLLL
+ 2 \left|{\gLL}\right|^{2}\right)\, 
\eea
\bea
\beta^{(1)}(m^2_{S^q_R}) &=&  12 \lamLRq m^2_{S^q_L}
+m^2_{S^q_R}\left(
-  \frac{2}{5} (C_{\Y}^q)^2 g^{\prime 2} 
-  8 g_3^{2} 
- 6 \gD^{2} 
+ 16 \lamRRQ
+ 2 \left|{\gQR}\right|^{2}\right)\, 
\eea
\bea
\beta^{(1)}(m^2_{S^L_R}) &=& 4 \lamLRl m^2_{S^L_R}
+ m^2_{S^L_R}\left(
- \frac{18}{5} (C_{\Y}^{\LL})^2 g^{\prime 2} 
- 6 \gD^{2} 
+ 8 \lamRRL
+ 2 \left|{\gLR}\right|^{2}\right)\, ,
\eea
where again $q=\;\scriptstyle{U,D}$ and $\scriptstyle{L=E,N}$, and $m^{2}_{S^{q,L}_{L,R}}$ denote the mass parameters of the messenger fields.
Although quantum correction inevitably generate a splitting in the messenger mass spectrum, the effect can be safely disregarded in the assessment of vacuum stability.

\section{Vacuum stability analysis}
\label{sec:5}

We can proceed with the analysis of the SM vacuum stability, dividing the study into the two aforementioned scenarios that cover complementary cases. In more detail, we consider:  
\begin{itemize}
\item {\bf Scenario S1.} After setting all the dark Yukawa couplings to a common perturbative value $g_{\LR}\equiv g_{\LL}^q=g_{\LL}^{{\LL}}=g_{\RR}^q=g_{\RR}^{{\LL}}$ at the matching scale $\mu_{\rm mes}\sim {\cal O}(m)$,  we compute the trilinear coupling $\Lambda_S=\lambda_S v_S$ needed to match the top Yukawa coupling through Eq.~(\ref{yukawatop}). Since the required value is in tension with the unitarity bound for the ratio $\Lambda_S/m$, we separately discuss a possible UV complete theory where the bound is avoided. In this scenario, the vacuum stability analysis is extended up to the Planck scale $\LambdaUV=M_{\rm Pl}$ due to the absence of Landau poles at lower scales.
  
\item {\bf Scenario S2.}
We assume a trilinear coupling $\Lambda_S = 4 \pi m$, corresponding to the  maximum value allowed by perturbative unitarity at low energy, and a large value of the mixing parameter $\xi=0.95$. We then initialize the couplings  $g^q_{\LR}\equiv g_{\LL}^q=g_{\RR}^q$ and  $g_{\LR}^{\LL}\equiv g_{\LL}^{{\LL}}=g_{\RR}^{{\LL}}$ by matching the Yukawa couplings of top quark and tau lepton. In this case, the resulting value of $g^q_{\LR}$ is necessarily borderline with perturbation theory and inevitably causes the emergence of a Landau pole at a scale $M_{\rm {\rm LP}} \ll M_{\rm Pl}$. Therefore, in the present scenario we aim to ensure that vacuum stability is at least achieved for energies as large as $\LambdaUV=M_{\rm {\rm LP}}$, remarking that a complete assessment valid at arbitrarily large energies requires a dedicated study of the non-perturbative regime of the theory.
\end{itemize}

In both the cases, to assess the stability of the EW vacuum we analyze the running of the Higgs boson quartic coupling, using Eq.~(\ref{Veff}) which approximates well the full RG-improved potential. The vacuum stability is then ensured if $\lambda_H(\mu_H) >0$ for all values of the scale $\mu$ from the EW scale up to the UV cutoff $\LambdaUV$. Concerning the quartic couplings in the messenger sector, since they do not play a direct role in the vacuum stability analysis, we set them to a common perturbative value at the matching scale $\mu_{\rm mes}$. The coupling $\lamHHS$ in Eq.~\eqref{VHS} is set instead to vanish at the matching scale to minimize its contribution to the running of $\lambda_H$.
Concretely, we choose:
\bea
  \lamSX(\mu_{\rm mes}) = \lamLLX(\mu_{\rm mes})=\lamRRX(\mu_{\rm mes})=\lamLRX(\mu_{\rm mes})=0.1\,,\\ \nonumber 
 \lamHHS(\mu_{\rm mes})=0\, ,~~\lambda_{H_S}(\mu_{\rm mes})=0.1\, ,
\label{quartic-initial}
\eea
where the superscript $\scriptstyle{X=U,D,E,N}$. 

In order to avoid color and charge breaking minima, throughout the following analyses we require that all the mass parameters and quartic couplings of mediators be positive up to the scale $\LambdaUV$. In regard of this, the corresponding $\beta$-functions force the initial values of these parameters to be sizeable, although still well perturbative, in order to overcome the large negative contribution due to the $SU(3)_c$ gauge group.
   
\subsection{Scenario S1}
\label{sub:S1}

Following the standard approach, we approximate the effective SM Higgs potential with its tree-level form improved by the running coupling,
$V_{eff}(H)=\lambda_H(\mu) H^4/4$, and identify $\mu\sim H$. 

\begin{figure}[h]
  \centering
  \includegraphics[width=.49\textwidth]{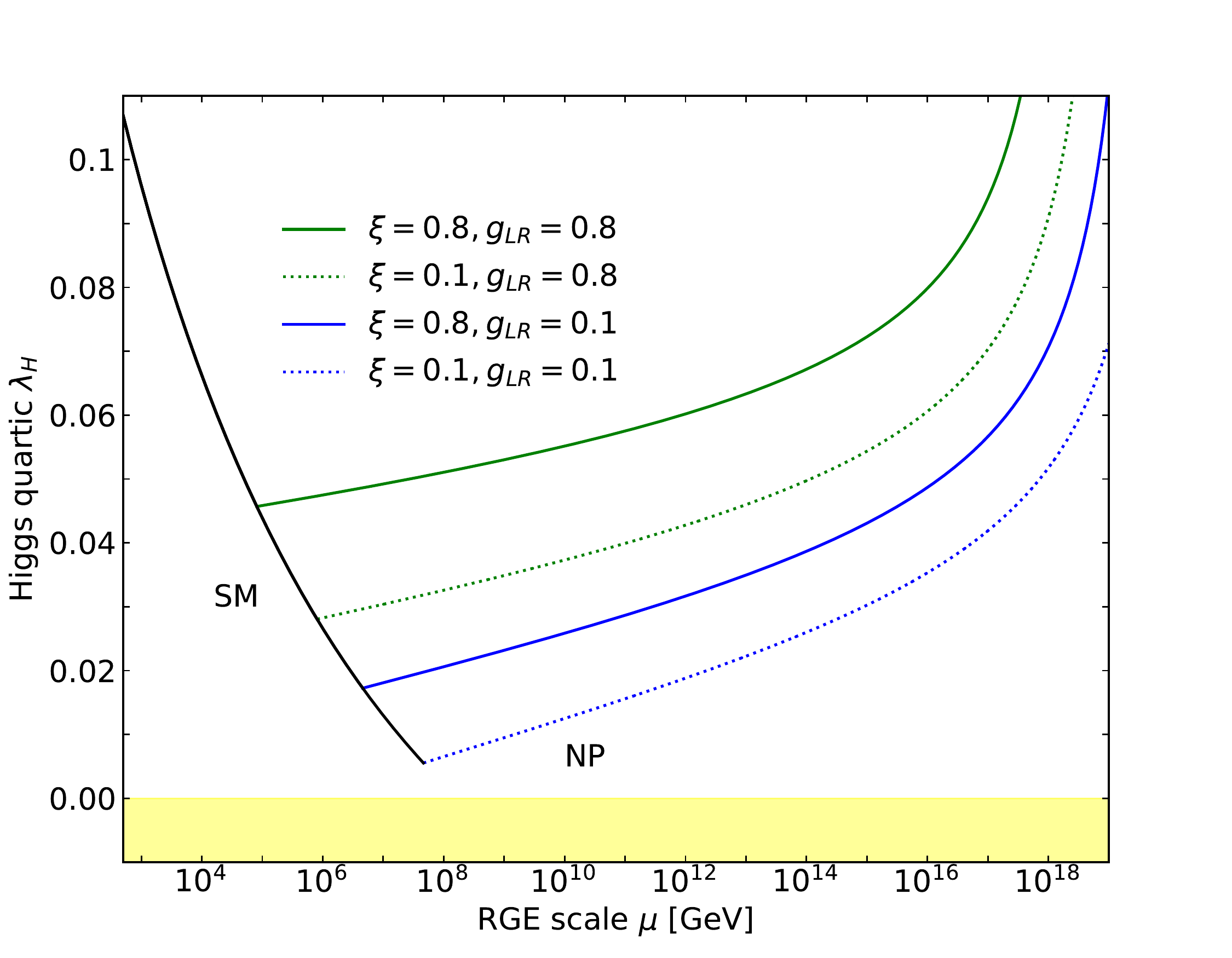}
  \includegraphics[width=.49\textwidth]{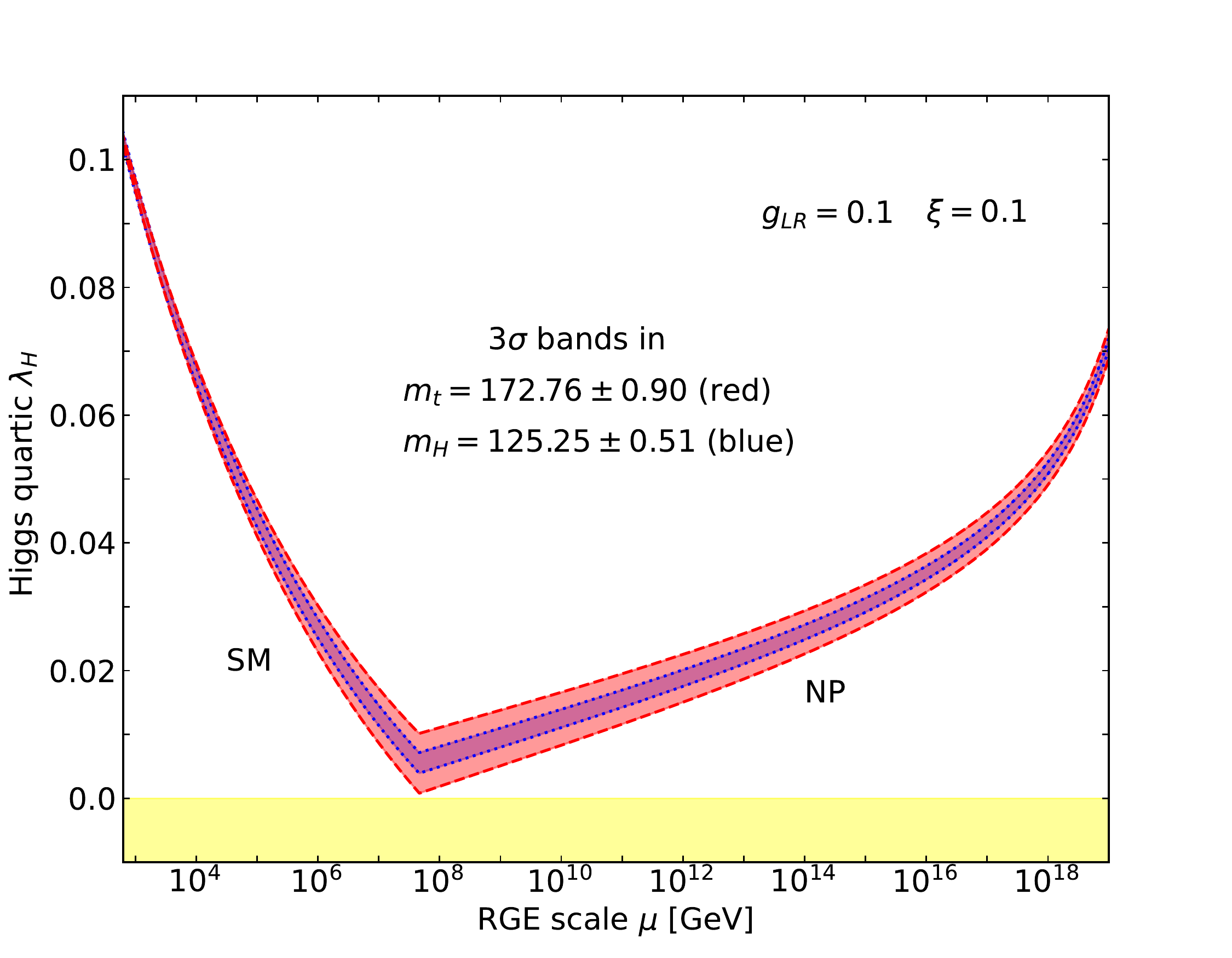}
  \caption{ Evolution of the effective Higgs boson quartic coupling $\lambda_H$ with the renormalization scale $\mu\sim H$,
  Eqs.~\eqref{eq:smlh} and (\ref{eq:fmh}). In the first panel, the colored lines represent the RG evolutions obtained for the indicated benchmark setups. In each case, the Higgs boson quartic coupling evolves according to the SM trajectory, shown in black, until it reaches the considered colored curve, then proceeds along the latter. The value of the renormalization scale at the point where the two curves meet corresponds to the matching scale. The second panel shows the effect of the current experimental uncertainties on one of the setups analyzed in the previous panel. In both the panels, the yellow band indicates the region excluded by vacuum stability.} 
    \label{fig2}
  \end{figure}

In Fig.~\ref{fig2} we present the results obtained by running the SM $\beta$-functions from the EW scale to the scale $\mu_{\rm mes}$, performing the matching and then running the parameters with the $\beta$-functions in Sec.~\ref{sec:4}. 

The first panel shows the evolution of the Higgs boson quartic coupling with the RGE scale, from $\mu=10^3$ GeV up to the UV cutoff identified in this scenario with the Planck scale. The yellow band signals the region of the parameter space ($\lambda_H(\mu)<0$) where the EW vacuum is \emph{not} stable.  

The different curves are obtained by considering the indicated combinations of the benchmark values used for the mixing parameter, $\xi=0.1, 0.8$, and the initial values of all dark Yukawa couplings, $g_{\LR}=0.1,0.8$. Once 
$g_{\LR}$ and  $\xi$ are selected, the value of the common messenger mass scale $m$ is set by the top Yukawa coupling in Eq.~(\ref{yukawatop}) through the matching conditions. As $m$ is also used as the matching scale, the full system of RGEs is solved iteratively until sufficient precision is obtained. 

In each of the analyzed cases, the matching scale corresponds to the value of the RG scale at which the running departs from the SM evolution, indicated by the black line. Correspondingly, at this point the top Yukawa coupling ceases to contribute to the $\lambda_H$ beta-function. The kinks in the curves are an artefact of the approximation used for the threshold conditions. For every curve, the ratio $\Lambda_S/m$ can be computed by inverting Eq.~\eqref{LambdaS-xi}: $\Lambda_S/m = \xi m /v$. For the values reported in the figure, in order of increasing matching scale $\mu_{\rm mes}\sim m$, we obtain $\Lambda_S/m= 2.6 \times 10^2, 3.3 \times 10^2,1.5 \times 10^5, 1.9 \times 10^5 $.

In the second panel of Fig.~\ref{fig2}, we show, instead, the effect of the current experimental uncertainties affecting the top quark and Higgs boson mass for the  previously analyzed case  that reaches closer to the instability zone. As we can see from these results, given the present measurements of these quantities, the stability of vacuum is always guaranteed in the present scenario. 

\subsection{Scenario S2}

\begin{figure}[h]
  \centering
  \includegraphics[width=.49\textwidth]{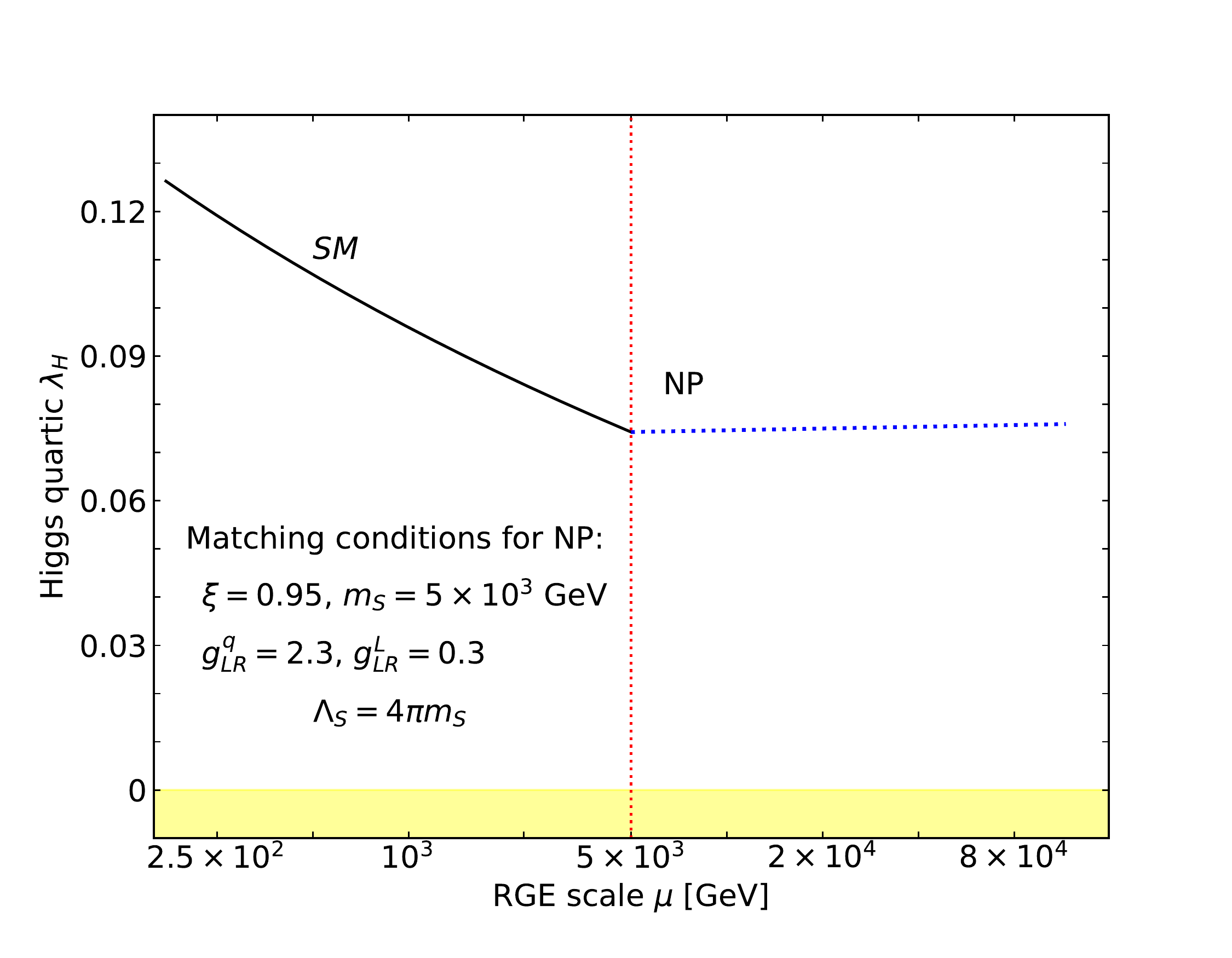}
  \includegraphics[width=.49\textwidth]{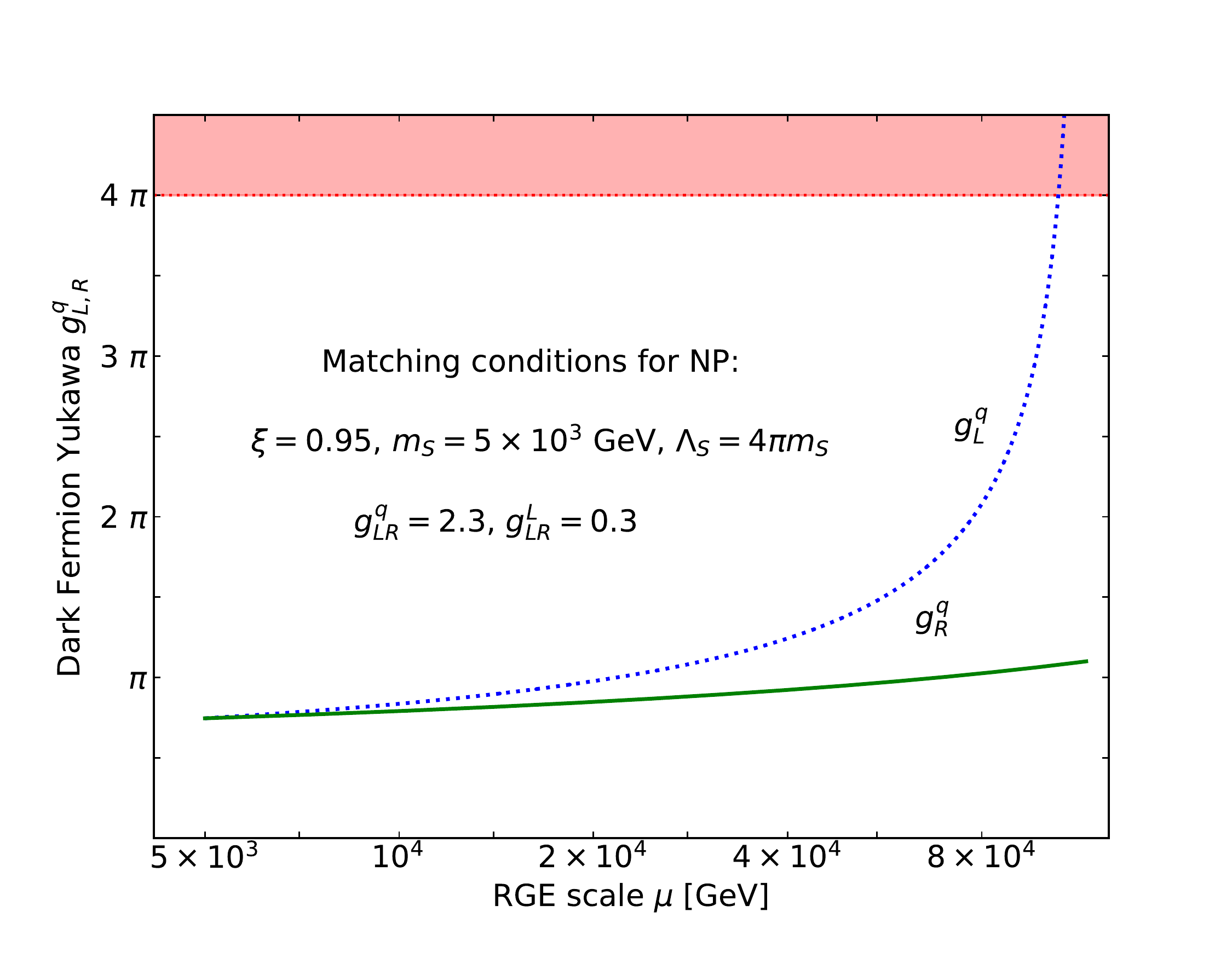}
  \caption{Left panel: evolution of the Higgs boson quartic coupling with the RG scale. The trilinear coupling $\Lambda_S$, entering the expressions obtained for the SM Yukawa operators, is set at the maximum value allowed by unitarity. The red dashed line indicates the matching scale, which separates the SM evolution of the parameter from its continuation determined by new physics. Right panel: evolution of the dark Yukawa coupling with the RG scale. Because of the large initial value selected by the considered value of $\Lambda_S$, the evolution of parameters is stopped by the emergence of a first Landau pole at $\mu\simeq10^5$ GeV, which defines the UV cutoff for the scenario. }
    \label{fig3}
  \end{figure}

The results obtained under the assumptions that specify the scenario S2 are shown in Fig.~\ref{fig3}.

 In more detail, the first panel shows again the RG evolution of the Higgs boson quartic coupling with the RG scale. In the plot, the trilinear coupling $\Lambda_S$ is set at the maximum value allowed by unitarity, which results in the initial conditions $g^q_{\LR}\simeq2.3$, $g^{\LL}_{\LR}\simeq0.3$ of the dark Yukawa couplings at the matching scale. The latter is denoted by the red dashed line and separates the SM evolution of $\lambda_H$ (black solid line) from its new physics continuation, rendered by the dotted segment in blue. Once again, the yellow band indicates the region of the parameter space where the vacuum is unstable. The RG evolution of the Higgs boson quartic coupling has been computed only up to $\mu\sim 10^5$ GeV, where the first Landau pole appears in the RG flow of dark Yukawa couplings related to quarks. This is illustrated in the second panel of Fig.~\ref{fig3}, which shows that the $g_{\LL}^q$ coupling is led to non-perturbative values (red band) already at a scale two order of magnitudes larger than the considered matching scale. The difference in the behaviors of $g_{\LL}^q$ and $g_{\RR}^q$ is solely due to the extra $SU(2)$ contributions in the $\beta$-function of the former.  

In Scenario S2, the EW vacuum stability is thus achieved at least in the energy range where the theory maintains perturbativity. The assessment of the vacuum structure of the theory for larger energies, however, requires non-perturbative methods that go beyond the scope of the present paper.

\subsection{Large trilinear couplings and unitarity bounds}
In this section we discuss the unitarity bound violated by the large trilinear coupling $\Lambda_s/m  \gg 4\pi$ of scenario S1 and speculate on a UV scenario where the constraint is relaxed.

As is well known, trilinear scalar interactions are UV-safe because the corresponding dimension 3 operator guarantees that processes mediated by these interactions do not violate unitarity in the limit of high energy, regardless of the value of the trilinear scalar coupling. On the other hand, at a set energy scale, the same interactions spoil unitarity when the trilinear coupling is much larger than any of the masses associated to the fields entering the trilinear vertex, as shown for instance in~\cite{DiLuzio:2016sur}.

For the case of scenario S1, the problematic trilinear coupling is due to the interaction vertex between the messenger fields and the Higgs boson $H$, arising after the $H_S$ field acquires a VEV: $\mathcal{L}\supset\Lambda_S \sum_i S_i S^{\dag}_i H$ -- see Eq.~(\ref{Vtrilinear}). We can then consider the scattering process 
\bea
S_i S_i \to S_j S_j
\label{SSscattering}
\eea
with $i\neq j$ allowed by the interaction in Eq.~(\ref{Vtrilinear}). The only diagram contributing to  the amplitude has a SM Higgs boson propagator in the s-channel. The amplitude $\mathcal{M}$ is given by
\bea
\mathcal{M}(s)=i\frac{\Lambda_S^2}{s-m_H^2}\, ,
\label{amplitude-SS}
\eea
where $s=(p_1+p_2)^2$ is the square of the center of mass energy and $p_1$ and $p_2$ the four-momenta of the initial state messengers $S_i$.

For center of mass energies comparable with the messengers mass threshold,  $s\simeq 4 m^2$,  the amplitude tends to 
\bea
\label{eq:ampthr}
\mathcal{M}(s\to 4 m^2) \to i\frac{\Lambda_S^2}{4m^2}\, ,
\eea
where we assumed $m\gg m_H$ and neglected the contribution of the Higgs mass in the denominator. As we can see, the amplitude of the process grows arbitrarily for $\Lambda_S/m\gg 1$, breaking the $S$-matrix unitarity at any fixed value of the scattering energy $s$. In particular, one can also show that for $\Lambda_S > 4\pi m$ perturbative unitarity is broken~\cite{DiLuzio:2016sur}. In fact, $\Lambda_S/m$ effectively works as a dimensionless coupling and consequently the ratio cannot be arbitrarily large if perturbation theory is to work. Still, for any fixed value of the trilinear coupling, the cross section at large energies $s\gg m^2$ scales as $1/s$ and thus the unitarity problem appears only at a set energy scale. Similar conclusions apply to the case of elastic scattering, where additional $t$- and $u$-channels diagrams contribute to the amplitude.

In order to recover perturbative unitarity, we explore an extension of the framework that adds a Lee-Wick (LW) higher derivative term~\cite{Lee:1971ix,Lee:1970iw} for the SM Higgs boson in the Lagrangian. The resulting kinetic term, which contains a fourth derivative of the field, can be rewritten as a sum of the usual scalar propagator plus the propagator of an unstable massive particle with negative norm: the LW ghost. The instability of the LW ghost formally allows to recover the unitarity of the  $S$ matrix upon restricting the asymptotic (stable) states of Hilbert space to positive norm states~\cite{Lee:1971ix,Lee:1970iw,Donoghue:2021eto}. The LW extension of the SM has been previously proposed in the context of the hierarchy problem related to the Higgs boson mass~\cite{Grinstein:2007mp,Espinosa:2011js,Carone:2008bs}.

Extending our model to include a LW higher derivative term for the Higgs field, the amplitude in Eq.~(\ref{amplitude-SS}) is modified by the propagation of the associated LW ghost, of mass $M_H$, as follows\footnote{The imaginary contribution in the propagator is neglected since the latter is computed off-shell.}:
\bea
\mathcal{M}^{\rm LW}(s)=i\Lambda_S^2\left(\frac{1}{s-m_H^2} -\frac{1}{s-M_H^2}\right)\, .
\label{amplitude-LW}
\eea
Assuming now that the messenger mass is larger than the LW ghost scale,  $m\gg M_H$, the amplitude at the threshold in Eq.~\eqref{eq:ampthr} becomes
\bea
\mathcal{M}^{\rm LW}(s\to 4 m^2) \to -i\frac{\Lambda_S^2 M_H^2}{16 m^4}
+{\cal O}(M_H^2/m^2) \, .
\label{amplitude-LW-SS}
\eea
Then, in the LW modified theory, perturbative unitarity of the process at hand is always guaranteed if $\frac{\Lambda_S^2 M_H^2}{m^4} \lsim 1$, implying
\bea
m_H \ll M_H < \frac{m^2}{\Lambda_S}\, .
\label{boundLW}
\eea
Interestingly, for the typical values used during the analysis of scenario S1, the characteristic scale of LW ghost is of order $\mathcal{O}(1-10)$ TeV, consistently with the lower bounds on the scenario from the LHC~\cite{Carone:2009nu}. In the present context, this indicates the expected cutoff that softens the SM Higgs hierarchy problem in the proposed LW extension.

As for the problem of EW vacuum stability, the presence of a higher derivative kinetic term for the Higgs boson modifies the RGEs of Section~\ref{sec:4}. In particular, the increased dependence of the propagator on inverse powers of the momentum makes all the one-loop diagrams presenting at least one Higgs boson circulating in the loop finite. Therefore, the corresponding contributions to the $\beta$-functions vanish above the mass scale associated to the LW ghost and the expressions in Section~\ref{sec:4} are modified as follows:

\bea
\label{eq:fmhLW}
\beta^{(1)}(\lambda_{H}) &=&
  \frac{1}{2} \lamHHS^{2}
- \frac{9}{5} g^{\prime 2} \lambda_{H}
- 9 g^{2} \lambda_{H}
+ \frac{27}{200} g^{\prime 4}
+ \frac{9}{20} g^{2} g^{\prime 2}
+ \frac{9}{8} g^{4}\, ,
\eea

\bea
\beta^{(1)}(\lambda_{H_S}) &=&
18 \lambda_{H_S}^{2}\, ,
\eea

\bea
\beta^{(1)}(\lamHHS) &=&
4N_F \Big( 3 (\lamSU)^2+ 3 (\lamSD)^2+ (\lamSL)^2 + (\lamSN)^2\Big) 
+  \lamHHS\Big(
   6 \lambda_{H_S} \nonumber \\
&-&  \frac{9}{10} g^{\prime 2}
-  \frac{9}{2} g^{2} \Big)\, ,
\eea

\bea
\beta^{(1)}(\lamSq) &=&\lamSq\left(
 2 \lamLRq
-  C^{q}_{\SSS} g^{\prime 2} 
-  \frac{9}{2} g^{2} 
- 8  g_{3}^{2} 
- 6 \gD^{2} 
+ \left|\gQL\right|^2
+ \left|\gQR\right|^2\right)\, ,
\eea

\bea
\beta^{(1)}(\lamSl) &=&\lamSl\left(
 2 \lamLRl
-  C^{\LL}_{\SSS} g^{\prime 2} 
-  \frac{9}{2} g^{2} 
- 6 \gD^{2} 
+ \left|\gLL\right|^2
+ \left|\gLR\right|^2\right)\, ,
\eea
with the superscript $q=\;\scriptstyle{U,D}$, $\scriptstyle{L=E,N}$, $C^{\U}_{\SSS}=13/10$, $C^{\D}_{\SSS}=7/10$, $C^{\E}_{\SSS}=27/10$ and $C^{\N}_{\SSS}=9/10$. As before, the couplings $g^{\prime}$, $g$, $g_3$, and $\gD$ correspond to the gauge groups $U(1)_Y$, $SU(2)_L$, $SU(3)_c$ and $U(1)_D$ respectively, and $N_F=3$ is the number of SM generations or families.

\begin{figure}[h]
  \centering
  \includegraphics[width=.49\textwidth]{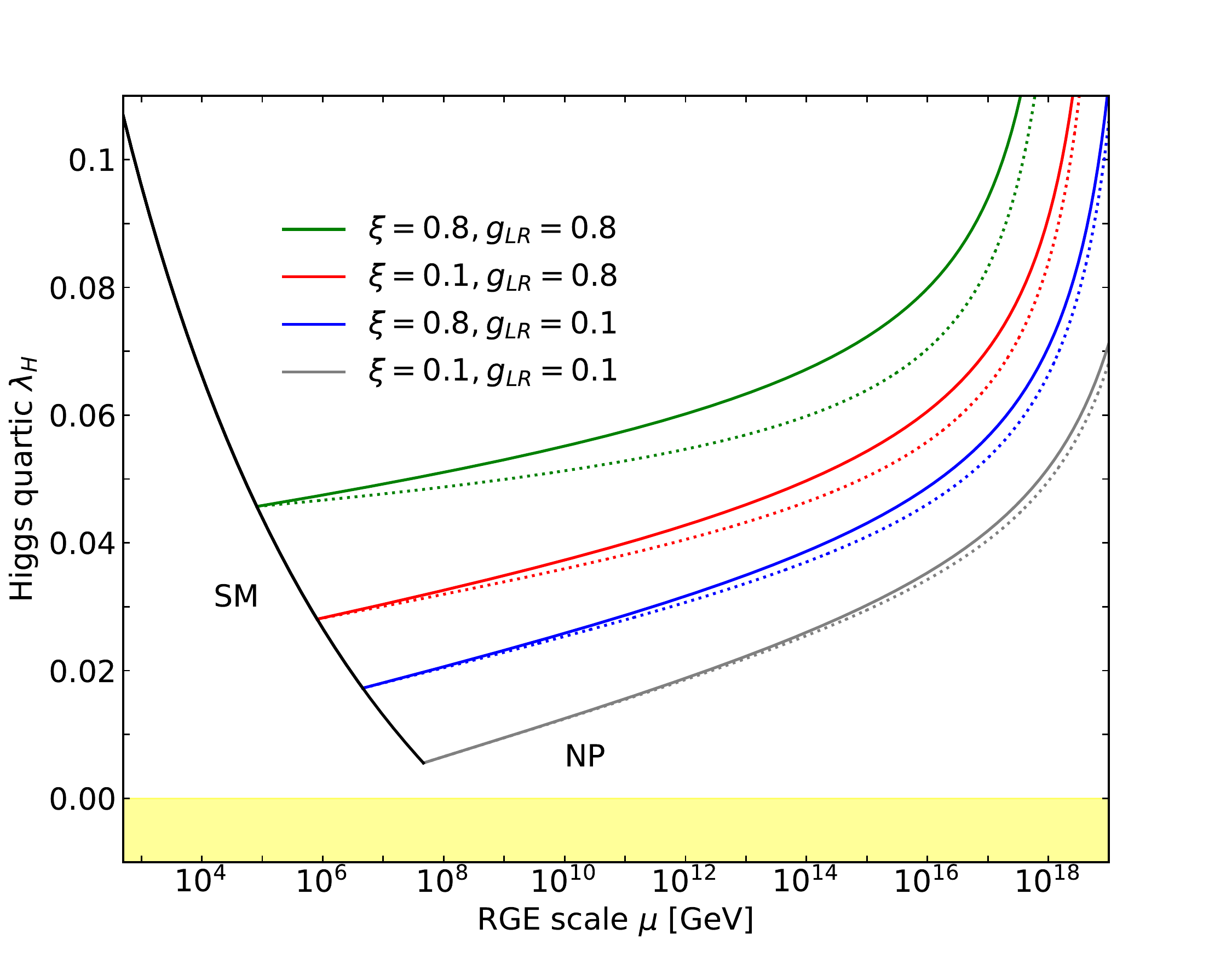}
  
  \caption{ Evolution of the effective Higgs boson quartic coupling $\lambda_H$ with the renormalization scale $\mu\sim H$  in the framework of Scenario S1 (solid lines) and in its LW extension (dotted lines) for the choice of parameters indicated in the legend.} 
    \label{fig:LW}
  \end{figure}

In order to assess the EW vacuum stability in the proposed LW extension, we have repeated the analysis of Scenario S1 using the above $\beta$-functions in place of the corresponding expressions presented in Section~\ref{sec:4}. In Figure~\ref{fig:LW} we compare the RG evolution of the Higgs boson quartic coupling obtained in the LW extension (dotted lines) to the results previously obtained (solid lines), shown also in the first panel of Figure~\ref{fig2}. In this example we have set the LW ghost mass scale to $M_H=1$ TeV, saturating the lower bound due to Eq.~\eqref{boundLW} to maximize the difference in the RG evolutions. As we can see, the Higgs boson quartic coupling remains positive on the whole range of values considered for the renormalization scale and its RG evolution qualitatively remains the same. Therefore, we conclude that the stability of EW vacuum can be guaranteed also in the proposed LW extension of the model.

\section{Conclusions}
\label{sec:6}

We have analysed the stability of the EW vacuum in the context of a previously proposed framework~\cite{Gabrielli:2013jka,Gabrielli:2016vbb,Gabrielli:2019sjg} for the radiative generation of the SM Yukawa interactions.  

In the simpler version~\cite{Gabrielli:2013jka} adopted in this paper, the framework uses a new discrete symmetry to first forbid the usual SM dimension 4 Yukawa operators. Then, non-perturbative effects related to a new $U(1)_{\rm D}$ gauge interaction yield a strongly hierarchical mass spectrum for a set of fermions charged under the symmetry. These dark fermions are in a one-to-one correspondence with the SM (Dirac) fermions and are responsible for sourcing the chiral symmetry breaking necessary to produce the SM Yukawa operators. The dark fermions and the SM particles are connected by a mediator sector, which hosts a set of scalar fields in a one-to-one correspondence with the SM Weyl fermions. Because the dark fermions are only charged under the $U(1)_{\rm D}$ gauge group, the messengers necessarily carry the same quantum numbers as squarks and sleptons of supersymmetric theories. The SM Yukawa couplings are thus generated at the one-loop level through processes allowed by the interactions of mediators, such as the one shown in Fig.~\ref{fig1}, after the spontaneous breaking of the discrete symmetry.

It is a peculiarity of the framework that the SM Higgs boson is naturally prevented from interacting directly with both the SM or the dark fermions, \emph{i.e.} that it is \emph{fermiophobic}. As a consequence, at energies higher than the mediator and dark fermion mass scales, fermions cannot contribute to the running of the Higgs boson Lagrangian parameters, in particular to its quartic coupling. 

Because the top quark Yukawa coupling provides the main contribution towards the metastability of the EW vacuum in the SM, in the present paper we set out to analyze the same problem of stability in light of the possible fermiophobic nature of the Higgs boson.

After detailing the model in Sec.~\ref{sec:Radiative Yukawa couplings} and reviewing the radiative generation of Yukawa couplings in Sec.~\ref{sec:3}, we show the RGEs for the full model in Sec.~\ref{sec:4}. In order to study the stability of the EW vacuum, we delineate two complementary scenarios that exemplify well the reach of the considered framework. The common strategy is to solve the SM RGEs up to a matching scale, identified with the messenger scale, and then evolve the parameters according to the RGEs of the full model.

In the first scenario, we study different cases were all the interactions between the SM fermions, the dark fermions and messenger fields are set to common and well perturbative benchmark values. Matching the top quark Yukawa coupling then requires the trilinear coupling appearing in the relevant amplitude, Eq.~\eqref{Ytopapprox}, to violate the unitarity bound of the $S$-matrix at energies close or below the matching scale. Postponing this issue momentarily, the results in Fig.~\ref{fig2} show that stability of EW vacuum can be achieved in the considered framework regardless of the current experimental uncertainties affecting the Higgs boson or the top quark mass. 
We argue that the unitarity of the $S$-matrix can be recovered in a LW extension of the framework that adds a LW ghost partner for the Higgs boson along the lines of the SM extension proposed in Refs.~\cite{Lee:1971ix,Lee:1970iw}. In this case, the model predicts a LW mass scale below 10 TeV, as required to solve the naturalness problem affecting the Higgs boson mass scale.

In the second scenario, instead, we take the maximal value of the trilinear coupling allowed by unitarity and initialize the relevant new physics interaction by matching the SM Yukawa couplings. As is evident from Eq.~\eqref{Ytopapprox}, the top quark case requires coupling with values that are borderline with perturbation theory. The stability analysis is then performed up to the scale of the first Landau pole emerging in the RGEs of the full model, identified here as an effective UV cutoff. The results shown in Fig.~\ref{fig3} show that the EW vacuum stability is ensured also in this case. 

Because the issue of vacuum stability does not significantly depend on the symmetry used to forbid the SM Yukawa couplings, the results obtained can be straightforwardly extended to the LR model presented in Ref.~\cite{Gabrielli:2016vbb}.

In conclusion, our analyses shows that the fermiophobic nature of the Higgs boson, imposed by the symmetries, can ensure the stability of the EW vacuum, regardless of the precise value of the top quark mass. The framework remarkably predicts the existence of weakly coupled dark sector fields, as well as of new messenger scalar interactions, that can be explored in the next generation of experiments at the LHC and future colliders.

\section*{Acknowledgements} 
\label{sec:Acknowledgements}
The authors thank Carlo Marzo and Martti Raidal for useful discussion. The work was supported by the European Union through the ERDF CoE grant TK133 and by the Estonian Research Council  through the grants PRG356 and MOBTT86. EG is affiliated to the Institute for Fundamental Physics of the Universe, Trieste, Italy.


\end{document}